\documentclass[aps,pra,preprint,superscriptaddress,showpacs]{revtex4}
\usepackage{bm}
\usepackage{amsmath}
\usepackage{dcolumn}
\newcolumntype{.}{D{x}{}{-1}}

\newcommand*{\centt}[1]{\multicolumn{1}{c}{#1}}
\newcolumntype{w}[1]{D{.}{.}{#1}}

\newcommand{\bsigma}{\vec{\sigma}}

\newcommand{\bfr}{\vec{r}}

\usepackage{graphicx}
\usepackage{xcolor}

\def\mynote1#1{{\color{blue}{\textsc{\footnotesize Question/Comment:} #1}}}
\def\mynote2#1{{\color{magenta}{#1}}}

\begin{document}
\preprint{Version 1.0}

\title{Higher-order recoil corrections for triplet states of the helium atom}
\author{Vojt\v{e}ch Patk\'o\v{s}}
\affiliation{Faculty of Physics, University of Warsaw,
             Pasteura 5, 02-093 Warsaw, Poland}

\author{V. A. Yerokhin}
\affiliation{Center for Advanced Studies, Peter the Great St.~Petersburg Polytechnic University,
195251 St.~Petersburg, Russia}

\author{Krzysztof Pachucki}
\affiliation{Faculty of Physics, University of Warsaw,
             Pasteura 5, 02-093 Warsaw, Poland}

\date{\today}

\begin{abstract}
Nuclear recoil corrections of order $\alpha^6\,m^2/M$ are calculated for the lowest-lying triplet
states of the helium atom. It improves the theoretical prediction for the isotope shift of
the $2^3S-2^3P$ transition energy and influences the determination of the
${}^3\textrm{He}-{}^4\textrm{He}$ nuclear charge radii difference. This calculation is a step
forward on the way towards the direct determination of the charge radius of the helium
nucleus from spectroscopic measurements.
\end{abstract}

\maketitle
%\tableofcontents

\section{Introduction}

The direct determination of the nuclear charge radius from the measured transition energies has
been so far carried out only for the hydrogen-like atoms \cite{mohr:12:codata}. In more complex
systems, the possibilities of such determination are  limited by our insufficient knowledge of the QED
effects. The main advantage of hydrogen-like atoms is that the relativistic electron wave
function can be determined analytically in the limit of infinite nuclear mass. It is then possible
to express all QED and nuclear recoil corrections within the Furry picture of QED and calculate
them either analytically in terms of the $Z\,\alpha$ expansion or numerically to all orders in
$Z\,\alpha$ (where $Z$ is the nuclear charge number and $\alpha$ is the fine structure constant).

Calculations of QED effects in few-electron systems are much more difficult than in hydrogen.
Presently the best theoretical accuracy is achieved for the helium atom, whose (low-lying) energy
levels are calculated rigorously within QED up to orders $\alpha^6\,m$ and $\alpha^5\,m^2/M $
\cite{pachucki:06hesinglet,yerokhin:10:helike} (where $m$ is the electron mass and $M$ is the
nuclear mass). The theoretical accuracy achieved in these calculations was not sufficient for
determination of the charge radius of the helium nucleus (i.e. $\alpha$-particle). Significant progress, however, can be achieved by calculating the next-order QED and nuclear recoil
effects, namely $\alpha^6\,m^2/M$ and $\alpha^7\,m$ corrections. These calculations will bring the
theoretical accuracy of the helium $n=2$ transition energies on a 10~kHz level, which will allow us
to determine the $\alpha$-particle charge radius with an accuracy of a few parts of $10^{-3}$. Such a
project is challenging but looks feasible, at least for the triplet states.

The most suitable transition for such a project is $2^3S - 2^3P$, which has already been
measured with sufficient accuracy \cite{cancio,hefs},
\begin{equation}
E(2^3{\rm S} - 2^3{\rm P}, ^4{\rm He})_{\rm centroid} =  276\,736 \,495\,649.5(2.1)\;{\rm kHz\,h}. \label{01}
\end{equation}
The finite nuclear size contribution to this transition energy is $E_{\rm fs} = 3\,427$ kHz\,h.
Taking into account that $E_{\rm fs}$ is proportional to the nuclear charge radius squared, $R^2$,
the expected 10-kHz theoretical accuracy will determine the nuclear charge radius with $0.15$\%
accuracy,
\begin{equation}\label{02}
\frac{\Delta R}{R} = \frac{1}{2}\,\frac{\delta E_{\rm fs}}{E_{\rm fs}}
\approx \frac{1}{2}\,\frac{10}{3\,427} \approx 1.5\cdot10^{-3}\,.
\end{equation}
After the project is accomplished, we shall be able to compare the charge radius of the
$\alpha$-particle with the result from muonic helium, which is expected soon from the CREMA
collaboration \cite{muHe}. Such a comparison would be of particular interest in view of the
discrepancy for the proton charge radius observed in the muonic hydrogen experiment
\cite{pohl:10,antognini:13}. Additional motivations for this project are the yet unexplained 
$4\,\sigma$ discrepancy for the difference in $^3$He and $^4$He nuclear charge radii
\cite{pachucki:15:heis} and plans to measure the charge radii difference from isotope shifts in
helium-like ions \cite{helikeIS}. In this work we make the first step towards 
the absolute nuclear charge radius determination
and calculate the nuclear recoil correction to order $\alpha^6\,m^2/M$ for the
$2^3S$ and $2^3P$ states of the helium atom.

This paper is organized as follows. Section II introduces notations that will be used throughout
the paper. Section III describes our approach to the calculation of the energy levels by an expansion
in the fine-structure constant $\alpha$. Section IV reports the Foldy-Wouthuysen transformed NRQED
Lagrangian, which is the starting point for our derivation. The derivation of the $\alpha^6\,m^2/M$
correction is presented in Section V. Section VI is devoted to the rearrangements of terms in such
a way that all matrix elements become finite. Section VII presents the final formulas. Section VIII
describes the numerical evaluation of all matrix elements. Results and discussion are presented in
Section IX. The principles of the dimensional regularization, details about the elimination of
singularities, the simplification of the formulas, and the reduction to the hydrogenic limit are
presented in Appendices.

\section{Notations}

We will use the following notations throughout the paper. The operators, energies, and wave
functions for a nucleus with a finite mass $M$ will be marked with indices ``$M$'': $X_M$, $E_M$,
$\phi_M$. The operators, energies, and wave functions in the infinite nuclear mass limit are
without indices: $X$, $E$, $\phi$.
The recoil corrections to the operators and energies are denoted by $\delta_M X$ and $\delta_M E$,
\begin{equation}
X_M\equiv X +\frac{m}{M}\delta_M X+O\Bigl(\frac{m}{M}\Bigr)^2\,,
\end{equation}
\begin{eqnarray}
E_M=E+\frac{m}{M}\,\delta_M E+O\Bigl(\frac{m}{M}\Bigr)^2\,.
\end{eqnarray}
We also introduce the shorthand notations:
\begin{equation}\label{09}
\langle X\rangle_M\equiv\langle\phi_M|X|\phi_M\rangle\,,
\end{equation}
and
\begin{eqnarray}\label{10}
\delta_M\,\langle X\rangle
&\equiv&
\biggl\langle\phi\,\biggl|\frac{ \vec{P}_I^2}{2}\frac{1}{(E-H)'}\,X\,\biggr|\,\phi\biggr\rangle
+\biggl\langle\phi\,\biggl|\, X\,\frac{1}{(E-H)'}\,\frac{ \vec{P}_I^2}{2}\biggr|\,\phi\biggr\rangle\,,
\end{eqnarray}
where $\vec{P_I}$ is the momentum of the nucleus in the center of mass frame, and 
$H$, $E$, and $\phi$  are the nonrelativistic
Hamiltonian, energy, and the wave function in the infinite nuclear mass limit.

\section{NRQED approach}
According to QED theory, the expansion of energy levels in powers of $\alpha$ has the form
\begin{eqnarray}\label{03}
E_M(\alpha)=E_M^{(2)}+E_M^{(4)}+E_M^{(5)}+E_M^{(6)}+E_M^{(7)}+O(\alpha^8),
\end{eqnarray}
where $E_M(\alpha)\equiv E(\alpha,\frac{m}{M})$, $E_M^{(n)}$ is a contribution of order $m\,\alpha^n$ and may include powers of $\ln\alpha$.
$E_M^{(n)}$ is in turn expanded in powers of the electron-to-nucleus mass ratio $m/M$
\begin{eqnarray}\label{04}
E_M^{(n)}=E^{(n)}+\frac{m}{M}\,\delta_M E^{(n)}+O\Bigl(\frac{m}{M}\Bigr)^2.
\end{eqnarray}
Each term of the expansion $E_M^{(n)}$ can be expressed as an expectation value of some effective
operator. Namely, $E_M^{(2)}\equiv E_M$ is the eigenenergy of the nonrelativistic Hamiltonian
$H_M^{(2)} \equiv H_M$ with the eigenstate $\phi_M$
\begin{eqnarray}\label{06}
H_M=\sum_a\left(\frac{\vec{p_a}^2}{2m}-\frac{Z\alpha}{r_{aI}}\right)+\sum_{a>b}\sum_{b}\frac{\alpha}{r_{ab}}+\frac{\vec{P_I}^2}{2M}.
\end{eqnarray}
Here $\vec{P_I}$ is the momentum of the nucleus; in the center of mass system it is just
$\vec{P_I}=-\sum_a \vec{p}_a$. $E_M^{(4)}$ is the expectation value of the Breit-Pauli Hamiltonian
$H_M^{(4)}$ \cite{bs},
\begin{equation}
E_M^{(4)} = \bigl\langle H_M^{(4)} \bigr\rangle_M\,,
\end{equation}
\begin{align} \label{07} H_M^{(4)} &\ =\sum_a \biggl[-\frac{\vec p^{\,4}_a}{8\,m^3} +
  \frac{ \pi Z\,\alpha}{2\,m^2}\,\delta^3(r_{aI}) +\frac{Z\,\alpha}{4\,m^2}\, \vec\sigma_a\cdot\frac{\vec
    r_{aI}}{r_{aI}^3}\times \vec p_a\biggr]
  \nonumber \\
  & +\sum_{a<b}\biggl\{ -\frac{\pi\,\alpha}{m^2}\, \delta^3(r_{ab}) -\frac{\alpha}{2\,m^2}\, p_a^i\,
  \biggl(\frac{\delta^{ij}}{r_{ab}}+\frac{r^i_{ab}\,r^j_{ab}}{r^3_{ab}}
  \biggr)\, p_b^j \nonumber \\ & - \frac{2 \pi\,\alpha}{3\,m^2}\,\vec\sigma_a
  \cdot\vec\sigma_b\,\delta^3(r_{ab}) +\frac{\alpha}{4\,m^2}\frac{\sigma_a^i\,\sigma_b^j} {
    r_{ab}^3}\,
  \biggl(\delta^{ij}-3\,\frac{r_{ab}^i\,r_{ab}^j}{r_{ab}^2}\biggr) \nonumber
  \\ & +\frac{\alpha}{4\,m^2\,r_{ab}^3} \bigl[ 2\,\bigl(\vec\sigma_a\cdot\vec
  r_{ab}\times\vec p_b - \vec\sigma_b\cdot\vec r_{ab}\times\vec p_a\bigr)
  \nonumber \\ & + \bigl(\vec\sigma_b\cdot\vec r_{ab}\times\vec p_b -
  \vec\sigma_a\cdot\vec r_{ab}\times\vec p_a\bigr)\bigr]\biggr\} \nonumber \\
 & - \frac{Z\,\alpha}{2\,m\,M}\, \sum_{a}
  \biggl[\frac{\bfr_{aI}}{r_{aI}^3}\times \vec{P_I} \cdot\bsigma_a -
  p_a^i\,\left(\frac{\delta^{ij}}{r_{aI}}+\frac{r^i_{aI}r^j_{aI}}{r_{aI}^3}\right) P_I^j
  \biggr] \,,
\end{align}
$E^{(5)}_M$ is the leading QED correction (see, e.g.,
\cite{helamb, sy, simple}), which will not be needed in the present
investigation. The next expansion term $E^{(6)}_M$ is the sum of two parts,
\begin{eqnarray}\label{08}
E_M^{(6)}=\left\langle H_M^{(4)}\,\frac{1}{(E_M-H_M)'}\,H_M^{(4)}\right\rangle_M + \bigl\langle H_M^{(6)}\bigr\rangle_M\,,
\end{eqnarray}
In this paper we derive the recoil part of this expression, $\delta_M E^{(6)}$, for triplet states
in helium and helium-like ions. The approach is similar to that in Ref. \cite{fw, pachucki:06hesinglet}, 
with some modifications that simplify the derivation of the recoil correction.

\section{Foldy-Wouthuysen transformation}
In order to derive the effective Hamiltonians $H_M^{(n)}$, and in particular $H_M^{(6)}$, we
transform the QED Lagrangian to the NRQED form by using the Foldy-Wouthuysen (FW) transformation \cite{itzykson}.
This transformation is the nonrelativistic expansion of the Dirac Hamiltonian in an external
electromagnetic field,
\begin{equation}
H = \vec \alpha \cdot \vec \pi +\beta\,m + e\,A^0\,, \label{11}
\end{equation}
where $\vec \pi = \vec p-e\,\vec A$. The FW
transformation $S$ 
\begin{equation}
H_{FW} = e^{i\,S}\,(H-i\,\partial_t)\,e^{-i\,S} = H + \delta H\,, \label{12}
\end{equation}
leads to a new Hamiltonian, which decouples the upper and lower components of the Dirac wave
function up to a specified order in the $1/m$ expansion.
In order to simplify the derivation of $m^2/M\,\alpha^6$ corrections,
we start from FW Hamiltonian from Ref. \cite{pachucki:06hesinglet},
\begin{eqnarray}\label{13}
 H_{FW} &=& e\,A^0 + \frac{\pi^2}{2\,m}-\frac{e}{4\,m}\,\sigma^{ij}\,B^{ij} -
\frac{\pi^4}{8\,m^3} + \frac{e}{16\,m^3}\bigl\{\sigma^{ij}\,B^{ij},\,p^2\bigr\}
\nonumber \\ &&
-\frac{e}{8\,m^2}\Bigl(\vec\nabla\cdot\vec E + \sigma^{ij}\,\bigl\{E^i,\,\pi^j\bigr\}\Bigr)
-\frac{e}{16\,m^3}\,\bigl\{\vec p\,,\,\partial_t\vec E\bigr\}
\nonumber \\ &&
+\frac{3\,e}{32\,m^4}\bigl\{\sigma^{ij}\,E^i\,p^j,\,p^2\bigr\}
+\frac{1}{128\,m^4}\,[p^2,[p^2,e\,A^0]]\nonumber \\ &&
-\frac{3}{64\,m^4}\,\Bigl(p^2\,\nabla^2 (e\,A^0) + \nabla^2 (e\,A^0)\,
p^2\Bigr) +\frac{1}{16\,m^5}\,p^6 ,
\end{eqnarray}
where $\{x\,,\,y\}$ and $[x,\,y]$ stand for the anti-commutator and commutator, correspondingly,
\begin{eqnarray}
\sigma^{ij}&=&\frac{1}{2\,i}\,[\sigma^i,\,\sigma^j]\,,\\
B^{ij}&=&\partial^i\,A^j-\partial^j\,A^i\,,\\
E^i&=&-\nabla^i\,A^0-\partial_t\,A^i\,,
\end{eqnarray}
and apply further transformations. The first one
\begin{equation}\label{14}
S_1 = -\frac{e}{16\,m^3}\bigl\{\vec \pi\,,\,\vec E\bigr\}
\end{equation}
eliminates $\partial_t\vec E$ from $H_{FW}$,
\begin{equation}\label{15}
\delta_1 H \approx
\frac{e}{16\,m^3}\bigl\{\vec p\,,\,\partial_t\vec  E\bigr\}
+\frac{e}{8\,m^3}\,\vec E^2 + \frac{1}{32\,m^4}\,[p^2, [p^2, e\,A^0]]\,.
\end{equation}
The second one
\begin{equation}\label{16}
S_2 = \frac{e}{8\,m^2}\,\sigma^{ij}\,\{A^i,\,\pi^j\},
\end{equation}
eliminates the transverse part $\vec E_\perp = -\partial_t \vec A$,
\begin{equation}\label{17}
\delta_2 H \approx \frac{e}{8\,m^2}\,\sigma^{ij}\{E^i_\perp,\,\pi^j\}
-\frac{e}{4\,m^2}\,\sigma^{ij}A^i\,E^j
+\frac{i\,e}{16\,m^3}\,[\,\sigma^{ij}\{A^i,\,p^j\}\,,\,p^2]\,.
\end{equation}
The resulting new FW Hamiltonian is
\begin{eqnarray}\label{18}
 H_{FW} &=& e\,A^0 + \frac{\pi^2}{2\,m}-\frac{e}{4\,m}\,\sigma^{ij}\,B^{ij} -
\frac{\pi^4}{8\,m^3} + \frac{e}{16\,m^3}\bigl\{\sigma^{ij}\,B^{ij},\,p^2\bigr\}
\nonumber \\ &&
-\frac{e}{8\,m^2}\Bigl(\vec\nabla\cdot\vec E_\parallel + \sigma^{ij}
\,\bigl\{E^i_\parallel,\,p^j\bigr\}\Bigr)
+\frac{e^2}{2\,m^2}\,\sigma^{ij}\,E^i_\parallel\, A^j
\nonumber \\ &&
+\frac{i\,e}{16\,m^3}\,[\,\sigma^{ij}\,\{A^i,\,p^j\}\,,\,p^2]
+\frac{e^2}{8\,m^3}\,\vec E^2_\parallel
+\frac{3\,e}{32\,m^4}\,\{p^2\,,\,\sigma^{ij}\,E^i_\parallel\,p^j\}
\nonumber \\ &&
+\frac{5}{128\,m^4}\,[p^2,[p^2,e\,A^0]]
-\frac{3}{64\,m^4}\,\Bigl\{p^2\,,\,\nabla^2 (e\,A^0)\Bigr\}
+\frac{1}{16\,m^5}\,p^6\,,
\end{eqnarray}
where $\vec E_\parallel = -\vec\nabla A^0$.
Since we are interested here in the leading
$O(m/M)$ term, the nucleus can be treated nonrelativistically, so
\begin{equation}\label{19}
\delta_M H_{FW} = \frac{1}{2\,M}\,\bigl(\vec{ P_I}+Z\,e\,\vec A)^2.
\end{equation}

\section{The higher order Breit-Pauli Hamiltonian}
In this section we derive the effective operator $H_M^{(6)}$. The derivation is similar to that in
Ref. \cite{pachucki:06hesinglet}, including the use of the dimensional regularization. For the simplicity of the
presentation, all the derivations here will be performed in $d=3$, but in such a way that allows for
a straightforward (and unique) generalization to the $d=3-2\,\epsilon$ form. This generalization 
will be needed only for a few divergent terms, and details of the dimensional regularization
are presented in Appendix A.

Using the nomenclature described in Appendix A, we denote by $V$ the nonrelativistic interaction
potential
\begin{equation}\label{20}
V \equiv \sum_a -\frac{Z\,\alpha}{r_{aI}} + \sum_{a>b}\,\sum_b\frac{\alpha}{r_{ab}}\,,
\end{equation}
by ${\cal E}_a$ the static electric field at the position of particle $a$
\begin{equation}\label{21}
e\,\vec{\cal E}_a \equiv -\nabla_a V =
-Z\,\alpha\,\frac{\vec r_{aI}}{r_{aI}^3} +\sum_{b\neq a}\alpha\,\frac{\vec r_{ab}}{r_{ab}^3}\,,
\end{equation}
by $\vec {\cal A}_{a}$ the vector potential at the position of particle  $a$,
which is produced by all other particles
\begin{eqnarray}\label{22}
e\,{\cal A}^i_{a} \equiv \sum_{b\neq a} \biggl[\frac{\alpha}{2\,r_{ab}}
\biggl(\delta^{ij}+\frac{r_{ab}^i\,r_{ab}^j}{r_{ab}^2}\biggr)\,
\frac{p_b^j}{m} + \frac{\alpha}{2\,m}\,\sigma^{ki}_b\frac{
  r_{ab}^k}{r_{ab}^3}\biggr]
-\frac{Z\alpha}{2r_{aI}}\biggl(\delta^{ij}+\frac{r_{aI}^ir_{aI}^j}{r_{aI}^2}\biggr)
\frac{P_I^j}{M}
\,,
\end{eqnarray}
and by $\vec {\cal A}_{I}$ the vector potential at the position of nucleus,
which is produced by electrons
\begin{eqnarray}\label{23}
e\,{\cal A}^i_{I} \equiv \sum_{a} \frac{\alpha}{2\,r_{aI}}
\biggl(\delta^{ij}+\frac{r_{aI}^i\,r_{aI}^j}{r_{aI}^2}\biggr)\,
\frac{p_a^j}{m}\,.
\end{eqnarray}
Following Ref. \cite{fw}, $H_M^{(6)}$ is expressed as a sum of various contributions
\begin{equation}\label{24}
H_M^{(6)} =\sum_{i=1,11} H_i^M\,.
\end{equation}
$H_{1}^M$ is the  kinetic energy correction
\begin{equation}\label{25}
H_{1}^M = \sum_a\frac{p_a^6}{16\,m^5}\,.
\end{equation}
$H_{2}^M$ is a correction due to the static electric interaction, namely
\begin{eqnarray}\label{26}
H_{2}^M &=& \sum_a\biggl(
\frac{e^2}{8\,m^3}\,\vec {\cal E}_a^2
+\frac{3}{32\,m^4}\,\{p_a^2\,,\,e\, \sigma^{ij}_a\,{\cal E}^i_a\,p^j_a\}
\nonumber \\ &&
+\frac{5}{128\,m^4}\,[p_a^2,[p_a^2,V]]
-\frac{3}{64\,m^4}\,\Bigl\{p_a^2\,,\,\nabla_a^2 V\Bigr\}\biggr).
\end{eqnarray}
$H_{3}^M$ is a correction
to the Coulomb interaction between electrons,
which comes from the 6$^{\rm th}$ term in $H_{FW}$, namely
\begin{equation}
-\frac{e}{8\,m^2}\Bigl(\vec\nabla\cdot\vec E_\parallel + \sigma^{ij}
\bigl\{E^i_\parallel,\, p^j\bigr\}\Bigr).
\label{27}
\end{equation}
If the interaction of both electrons is modified by this term,
it can be obtained in the non-retardation approximation, so
\begin{eqnarray}
H_{3}^M &=& \sum_{a>b}\sum_b
\int d^3 k\,\frac{4\,\pi}{k^2}\,\frac{1}{64\,m^4}\,
\biggl(k^2 +2\,i\,\sigma^{ij}_a k^i p^j_a\biggr)\,
e^{i\,\vec k\cdot\vec r_{ab}}\,
\biggl(k^2 +2\,i\,\sigma^{kl}_b k^k p^l_b\biggr)
\nonumber \\ &=& \sum_{a>b}\sum_b
\frac{1}{64\,m^4}\,\biggl\{
-4\,\pi\,\nabla^2\,\delta^3(r_{ab})
-8\,\pi\,i\,\sigma^{ij}_a\, p^i_a\,\delta^3(r_{ab})\, p^j_a
-8\,\pi\,i\,\sigma^{ij}_b\, p^i_b\,\delta^3(r_{ab})\, p^j_b
\nonumber \\ &&
+4\,\sigma^{ki}_a\,p^k_a
\biggl[ \frac{\delta^{ij}}{3}\,4\,\pi\,\delta^3(r_{ab})+
\frac{1}{r_{ab}^3}\,\biggl(\delta^{ij}-3\,\frac{r_{ab}^i\,r_{ab}^j}{r_{ab}^2}
\biggr)\biggr] \sigma^{lj}_b\, p^l_b
\biggr\}\,.\label{28}
\end{eqnarray}
$H_{4}^M$ is the relativistic correction
due to transverse photon exchange
\begin{eqnarray}\label{29}
H_{4}^M &=& \sum_a -\frac{e}{8\,m^3}\,\bigl(\pi_a^4-\frac{e}{2}\,\bigl\{\sigma^{ij}_a\, B^{ij}_a
,\,p_a^2\bigr\}\bigr)\nonumber \\ &=&
\sum_a \frac{e}{8\,m^3}\,\bigl(
2\,\{p_a^2\,,\,\vec p_a\cdot\vec{\cal A}_a\} +
\{p_a^2\,,\,\sigma^{ij}_a\,\nabla^i_a\,{\cal A}^j_a\}\bigr).
\end{eqnarray}
$H_{5}^M$ comes from the remaining transverse photon exchange
\begin{equation}\label{30}
H_5^M = \sum_a \biggl(\frac{e^2}{2\,m^2}\,\sigma^{ij}_a\,{\cal E}^{i}_a\, {\cal A}^j_{a}
+\frac{i\,e}{16\,m^3}\,[\,\sigma^{ij}\,\{{\cal A}^i_a,\, p^j_a\}\,,\,p_a^2]\biggr)\,.
\end{equation}
$H_{6}^M$ comes from the double transverse photon exchange
\begin{equation}\label{31}
H_6^M = \sum_a\frac{e^2}{2\,m^2}\,{\cal A}_a^2  +
\frac{Z^2\,e^2}{2\,M}\,{\cal A}_I^2 .
\end{equation}
$H_7^M$ is a retardation correction in the nonrelativistic single
transverse photon exchange
\begin{eqnarray}\label{32}
E_7^M &=&  -e^2\,\int\frac{d^3k}{(2\,\pi)^3\,2\,k^4}\,
\biggl(\delta^{ij}-\frac{k^i\,k^j}{k^2}\biggr)\,\biggl[
\,\sum_{a \neq b}\sum_b
\biggl\langle\phi_M\,\biggl|
\biggl(\frac{ p^i_a}{m}+\frac{1}{2\,m}\,
\sigma^{ki}_a\,\nabla^k_a\biggr)
\,e^{i\,\vec k\cdot\vec r_a}\,\nonumber\\&&
(H_M-E_M)^2
\,\biggl(\frac{p^j_b}{m}+\frac{1}{2\,m}\,
\sigma^{lj}_b\,\nabla^l_b\biggr)
\,e^{-i\,\vec k\cdot\vec
r_b}\,\biggr|\phi_M\biggr\rangle
\nonumber \\ &&
-Z\,\sum_b\biggl\langle\phi\,\biggl|
\frac{ P_I^i}{M}\,e^{i\,\vec k\cdot\vec r_I}\,(H-E)^2
\,\biggl(\frac{p^j_b}{m}+\frac{1}{2\,m}\,
\sigma^{lj}_b\,\nabla^l_b\biggr)
\,e^{-i\,\vec k\cdot\vec
r_b}\,\biggr|\phi\biggr\rangle
\nonumber \\ &&
-Z\,\sum_a\biggl\langle\phi\,\biggl|
\biggl(\frac{p^i_a}{m}+\frac{1}{2\,m}\,
\sigma^{ki}_a\,\nabla^k_a\biggr)
\,e^{i\,\vec k\cdot\vec r_a}\,
(H-E)^2\,\frac{ P_I^j}{M}
\,e^{-i\,\vec k\cdot\vec
r_I}\,\biggr|\phi\biggr\rangle
\biggr]\,.
\end{eqnarray}
This is the most complicated term in the evaluation,
and we have to split it into four parts with no spin, single spin,
and double spin terms, and the nuclear part
\begin{equation}
E_7^M = E_{7a}^M + E_{7b}^M + E_{7c}^M + E_{7d}^M.\label{33}
\end{equation}
The part with double spin operators is
\begin{eqnarray}
E_{7c}^M &=& \sum_a \sum_{a\neq b}-e^2\int\frac{d^3k}{(2\,\pi)^3\,2\,k^4}
\frac{(\sigma^{ki}_a\, k^k)\,(\sigma^{li}_b\, k^l)}{4\,m^2}\,
\Bigl\langle\phi_M\Bigl|
\,e^{i\,\vec k\cdot\vec r_a}\,(H_M-E_M)^2
\,e^{-i\,\vec k\cdot\vec
r_b}\,\Bigr|\phi_M\Bigr\rangle.\label{34}\nonumber\\
\end{eqnarray}
One uses the commutation identity
\begin{eqnarray}
\Bigl\langle\,e^{i\,\vec k\cdot\vec r_a}\,(H_M-E_M)^2\,e^{-i\,\vec k\cdot\vec r_b}\Bigr\rangle_M
+(a\leftrightarrow b)&=&
\Bigl\langle\Bigl[e^{i\,\vec k\cdot\vec r_a},\Bigl[(H_M-E_M)^2, e^{-i\,\vec
      k\cdot\vec r_b}\Bigr]\Bigr]\Bigr\rangle_M
\nonumber \\ &=&
-\frac{1}{2\,m^2}\,\Bigl\langle
\bigl[p_a^2,\bigl[p_b^2,e^{i\,\vec k\cdot\vec r_{ab}}\bigr]\bigr]
\Bigr\rangle_M\label{35}
\end{eqnarray}
to express this correction in terms of the effective operator $H_{7c}^M$,
\begin{eqnarray}
H_{7c}^M &=& \sum_{a>b}\sum_b\frac{\alpha}{16\,m^4}\,
\biggl[p_a^2,\biggl[p_b^2,
    \sigma^{ij}_a\,\sigma^{ij}_b\,\frac{1}{3\,r_{ab}}+
\sigma_a^{i}\,\sigma_b^{j}\,\frac{1}{2\,r_{ab}}\biggl(\frac{r_{ab}^i\,r_{ab}^j}{r_{ab}^2}
-\frac{\delta^{ij}}{3}\biggr)\biggr]\biggr].\label{36}
\end{eqnarray}
The part with no spin operator is
\begin{eqnarray}
E_{7a}^M &=& \sum_{a\neq b}\sum_b -e^2\int\frac{d^3k}{(2\,\pi)^3\,2\,k^4}\,
\biggl(\delta^{ij}-\frac{k^i\,k^j}{k^2}\biggr)\,
\nonumber \\ &&
\biggl\langle\phi_M\biggl|\frac{p_a^i}{m}
\Bigl\{e^{i\,\vec k\cdot\vec r_a}\,(H_M-E_M)^2\,e^{-i\,\vec k\cdot\vec r_b}
-(H_M-E_M)^2\Bigr\}\,\frac{p_b^j}{m}\,
\biggr|\phi_M\biggr\rangle\,.\label{37}
\end{eqnarray}
We subtracted here the term with $k=0$. We ought to perform this
in Eq. (\ref{32}), but for simplicity of writing we have not done it until now.
We use another commutator identity
\begin{eqnarray}
&&e^{i\,\vec k\cdot\vec r_a}\,(H_M-E_M)^2\,e^{-i\,\vec k\cdot\vec r_b}
-(H_M-E_M)^2 =
\nonumber \\ &&
(H_M-E_M)\,(e^{i\,\vec k\cdot\vec r_{ab}}-1)\,(H_M-E_M)
+ (H_M-E_M)\,\biggl[\frac{p_b^2}{2\,m},e^{i\,\vec k\cdot\vec r_{ab}}-1\biggr]
\nonumber \\ && + \biggl[e^{i\,\vec k\cdot\vec r_{ab}}-1,\frac{p_a^2}{2\,m}\biggr]\,(H_M-E_M)
+ \biggl[\frac{p_b^2}{2\,m},\biggl[e^{i\,\vec k\cdot\vec
      r_{ab}}-1,\frac{p_a^2}{2\,m}\biggr]\biggr] \label{38}
\end{eqnarray}
and the integration formula
\begin{equation}
\int d^3k\frac{4\,\pi}{k^4}\,\biggl(\delta^{ij}-\frac{k^i\,k^j}{k^2}\biggr)\,
\bigl(e^{i\,\vec k\cdot\vec r}-1\bigr) = \frac{1}{8\,r}\,
\bigl(r^i\,r^j-3\,\delta^{ij}\,r^2\bigr)\label{39}
\end{equation}
to obtain the effective operator $H_{7a}^M$
\begin{eqnarray}
H_{7a}^M &=& \sum_{a>b}\sum_b -\frac{\alpha}{8\,m^2}\,\biggl\{
\bigl[p_a^i,V\bigr]\,\frac{r_{ab}^i\,r_{ab}^j-3\,\delta^{ij}\,r_{ab}^2}{r_{ab}}\,
\bigl[V,p_b^j\bigr]\nonumber \\ &&
+\bigl[p_a^i,V\bigr]\,\biggl[\frac{p_b^2}{2\,m},
 \frac{r_{ab}^i\,r_{ab}^j-3\,\delta^{ij}\,r_{ab}^2}{r_{ab}}\biggr]\,p_b^j
+p_a^i\,\biggl[\frac{r_{ab}^i\,r_{ab}^j-3\,\delta^{ij}\,r_{ab}^2}{r_{ab}},
\frac{p_a^2}{2\,m}\biggr]\,\bigl[V,p_b^j\bigr]
\nonumber \\ &&
+p_a^i\,\biggl[\frac{p_b^2}{2\,m},\biggl[
\frac{r_{ab}^i\,r_{ab}^j-3\,\delta^{ij}\,r_{ab}^2}{r_{ab}},
\frac{p_a^2}{2\,m}\biggr]\biggr]\, p_b^j
\biggr\}.\label{40}
\end{eqnarray}
The part with the single spin operator is
\begin{eqnarray}\label{41}
E_{7b}^M &=& \sum_{a\neq b}\sum_b-\frac{i\,e^2}{4\,m^2}
\int\frac{d^3 k}{(2\,\pi)^3\,k^4}
 \\ &&\hspace*{-5ex}
\Bigl\langle\phi_M\biggr|\Bigl\{e^{i\,\vec k\cdot\vec r_a}\,
(H_M-E_M)^2\,e^{-i\,\vec k\cdot\vec r_b}\,\sigma^{ki}_a\, k^k\, p^i_b
-\sigma^{lj}_b\, p^j_a\,k^l\,e^{i\,\vec k\cdot\vec r_a}\,
(H_M-E_M)^2\,e^{-i\,\vec k\cdot\vec r_b}\Bigr\}\biggr|\phi_M\Bigr\rangle. \nonumber
\end{eqnarray}
With the help of the commutator in Eq. (\ref{38}) and the integral
\begin{equation}
\int d^3k\,\frac{4\,\pi\,\vec k}{k^4}\,
e^{i\,\vec k\cdot\vec r} = \frac{i}{2}\,\frac{\vec r}{r}\label{42}
\end{equation}
one obtains
\begin{eqnarray}
H_{7b}^M &=& \sum_{a>b}\sum_b\frac{\alpha}{4\,m^2}\biggl\{
\biggl[\sigma^{ij}_a\frac{
    r^i_{ab}}{r_{ab}},\frac{p_a^2}{2\,m}\biggr]\,
    \bigl[V,p^j_b] + \biggl[\frac{p_b^2}{2\,m},
\biggl[\sigma^{ij}_a\frac{
    r^i_{ab}}{r_{ab}},\frac{p_a^2}{2\,m}\biggr]\biggr]\, p^j_b
\nonumber \\ &&
-\bigl[p^j_a,V\bigr]\,\biggl[p_b^2,  \sigma^{ij}_b\,\frac{
    r^i_{ab}}{r_{ab}}\biggr]- p^j_a\,\biggl[\frac{p_a^2}{2\,m},
\biggl[\sigma^{ij}_a\,\frac{
    r^i_{ab}}{r_{ab}}\,,\frac{p_b^2}{2\,m}\biggr]\biggr]\biggr\}.\label{43}
\end{eqnarray}
Finally, the nuclear part is
\begin{eqnarray}\label{44}
E_{7d}^M &=&  -e^2\,\int\frac{d^3k}{(2\,\pi)^3\,2\,k^4}\,
\biggl(\delta^{ij}-\frac{k^i\,k^j}{k^2}\biggr) \nonumber \\&&
\times\frac{Z}{M}\,\sum_{a,b}\biggl\langle\phi\,\biggl|
p_a^i\,(H-E)^2
\,\biggl(\frac{ p^j_b}{m}+\frac{1}{2\,m}\,
\sigma^{lj}_b\,\nabla^l_b\biggr)
\,e^{-i\,\vec k\cdot\vec r_{bI}}\,\biggr|\phi\biggr\rangle
+ {\rm h.c.}\nonumber\\&=&
-\frac{Z\,\alpha}{M}\sum_{a,b}\biggl\langle\phi\,\biggl| [p_a^i\,,\,V]\,\biggl[H-E\,,\,
\frac{(r_{bI}^i\,r_{bI}^j-3\,\delta^{ij}\,r_{bI}^2)}{8\,r_{bI}}\,\frac{p_b^j}{m}
-\frac{1}{4\,m}\,\sigma^{lj}_b\,\frac{ r^l_{bI}}{r_{bI}}\,
\biggr]\biggr|\phi\biggr\rangle\nonumber\\&\equiv&
\langle\phi|H_{7d}^M|\phi\rangle.
\end{eqnarray}
We have checked that the non-recoil part agrees with that derived in \cite{fw} and
that the spin-dependent recoil part agrees with that in \cite{fs_recoil}. Here, we are
interested in the spin-independent part, which in the center-of-mass system $\vec{P_I}=-\sum_a\vec{p}_a$ is
(from now on we use atomic units $m=1$)
\begin{eqnarray}
H_1^M &=&  \sum_a\frac{p_a^6}{16}\,,\nonumber \\
H_2^M &=& \sum_a\biggl(
\frac{(\nabla_a V)^2}{8}
+\frac{5}{128}\,[p_a^2,[p_a^2,V]]
-\frac{3}{64}\,\Bigl\{p_a^2\,,\,\nabla_a^2 V\Bigr\}\biggr)\,,\nonumber\\
H_{3}^M &=&  \sum_{a>b}\sum_b
\frac{1}{64}\,\biggl\{
-4\,\pi\,\nabla^2\,\delta^3(r_{ab})
+\frac{2}{3}\,\sigma^{ij}_a\,\sigma^{ij}_b\,\biggl[
\frac{2}{3}\,\vec p_a\,4\,\pi\,\delta^3(r_{ab})\,\vec p_b -
p_a^i\frac{1}{r_{ab}^3}\,\biggl(\delta^{ij}-3\,\frac{r_{ab}^i\,r_{ab}^j}{r_{ab}^2}
\biggr)\,p_b^j\biggr]\biggr\}\,,\nonumber\\
H_4^M &=& \frac{1}{8}\sum_a\biggl[\sum_{b\neq a}\biggl\{p_a^2\,,
\,p_a^i\biggl(\frac{\delta^{ij}}{r_{ab}} +
\frac{r_{ab}^i\,r_{ab}^j}{r_{ab}^3}\biggr)p_b^j\biggr\}
-\frac{\sigma^{ij}_a\,\sigma^{ij}_b}{6}\,\bigl\{p_a^2\,,\,4\,\pi\,\delta^3(r_{ab})\bigr\}
\nonumber \\&&
+\frac{Z}{M}\,\sum_b\biggl\{p_a^2\,,
\,p_a^i\biggl(\frac{\delta^{ij}}{r_{aI}} +
\frac{r_{aI}^i\,r_{aI}^j}{r_{aI}^3}\biggr)p_b^j\biggr\}
\biggr]\,,\nonumber\\
H_5^M &=&  \sum_{a\neq b,b}\frac{\sigma^{ij}_a\,\sigma^{ij}_b}{6}\,\biggl(
-\frac{1}{2}\,\frac{\vec r_{ab}}{r_{ab}^3}\cdot\nabla_a V +
\frac{1}{16}\,\biggl[\biggl[\frac{1}{r_{ab}}\,,\,p_a^2\biggr]\,,\,p_a^2\biggr]\biggr)\,,
\nonumber \\
H_6^M &=&  \sum_a\sum_{b\neq a}\sum_{c\neq a}\biggl[
\frac{1}{8}\,p_b^i\,\biggl(\frac{\delta^{ij}}{r_{ab}}+\frac{r_{ab}^i\,r_{ab}^j}{r_{ab}^3}\biggr)\,
\biggl(\frac{\delta^{jk}}{r_{ac}}+\frac{r_{ac}^j\,r_{ac}^k}{r_{ac}^3}\biggr)\,p_c^k
+\frac{\sigma^{ij}_b\,\sigma^{ij}_c}{24}\,
\frac{\vec r_{ab}}{r_{ab}^3}\,\frac{\vec r_{ac}}{r_{ac}^3}
\nonumber \\ &&
+\frac{Z}{4}\frac{m}{M}\,p_b^i\,\biggl(\frac{\delta^{ij}}{r_{ab}}+\frac{r_{ab}^i\,r_{ab}^j}{r_{ab}^3}\biggr)\,
\biggl(\frac{\delta^{jk}}{r_{aI}}+\frac{r_{aI}^j\,r_{aI}^k}{r_{aI}^3}\biggr)\,p_c^k\biggr]
+\sum_a\sum_b\frac{Z^2}{8}\,\frac{m}{M}\nonumber \\ &&
\times\biggl[p_a^i\,\biggl(\frac{\delta^{ij}}{r_{aI}}+\frac{r_{aI}^i\,r_{aI}^j}{r_{aI}^3}\biggr)\,
\biggl(\frac{\delta^{jk}}{r_{bI}}+\frac{r_{bI}^j\,r_{bI}^k}{r_{bI}^3}\biggr)\,p_b^k
+\frac{\sigma^{ij}_a\,\sigma^{ij}_b}{3}\,
\frac{\vec r_{aI}}{r_{aI}^3}\,\frac{\vec r_{bI}}{r_{bI}^3}\biggr]\,,
\nonumber \\
H_{7a}^M &=&\sum_{a>b}\sum_b -\frac{1}{8}\,\biggl\{
\bigl[p_a^i,V\bigr]\,\frac{r_{ab}^i\,r_{ab}^j-3\,\delta^{ij}\,r_{ab}^2}{r_{ab}}\,
\bigl[V,p_b^j\bigr]\nonumber \\ &&
+\bigl[p_a^i,V\bigr]\,\biggl[\frac{p_b^2}{2},
 \frac{r_{ab}^i\,r_{ab}^j-3\,\delta^{ij}\,r_{ab}^2}{r_{ab}}\biggr]\,p_b^j
+p_a^i\,\biggl[\frac{r_{ab}^i\,r_{ab}^j-3\,\delta^{ij}\,r_{ab}^2}{r_{ab}},
\frac{p_a^2}{2}\biggr]\,\bigl[V,p_b^j\bigr]
\nonumber \\ &&
+p_a^i\,\biggl[\frac{p_b^2}{2},\biggl[
\frac{r_{ab}^i\,r_{ab}^j-3\,\delta^{ij}\,r_{ab}^2}{r_{ab}},
\frac{p_a^2}{2}\biggr]\biggr]\, p_b^j\biggr\}\,,  \nonumber \\
H_{7c}^M &=&\sum_{a>b}\sum_b\frac{\sigma^{ij}_a\,\sigma^{ij}_b}{48}\,
\biggl[p_a^2,\biggl[p_b^2,\frac{1}{r_{ab}}\biggr]\biggr] \,, \nonumber \\
H_{7d}^M &=& \frac{i\,Z}{8}\frac{m}{M}\,\sum_{a,b}\nabla_a^i V\,\biggl[H-E\,,\,
\frac{(r_{bI}^i\,r_{bI}^j-3\,\delta^{ij}\,r_{bI}^2)}{r_{bI}}\,p_b^j\biggr]
\,.\label{46}
\end{eqnarray}
Further Hamiltonians $H^M_8\ldots H^M_{11}$ come from the high-energy contributions, so they are
proportional to Dirac delta's and we will account for them in the next paragraph. These $H^M_i$
form a general $m\,\alpha^6$ effective Hamiltonian for arbitrary atom and arbitrary state,
neglecting the spin-dependent operators.

From now on we consider the specific case of the triplet states of the He atom, where the expectation
value of $\delta^3(r_{ab})$ vanishes and almost all matrix elements become finite.

The Breit-Pauli Hamiltonian of Eq. (\ref{07}) is split into four parts (with $r_{12}\equiv r$,
$r_{aI}\equiv r_a$ and $\vec{P}\equiv\vec{p}_1+\vec{p}_2$)
\begin{equation}\label{47}
H_M^{(4)} = H_A^M+H_B^M+H_C^M+H_D^M\,,
\end{equation}
where
\begin{eqnarray}
H_A^M & = & -\frac{1}{8}\,(p_1^4+p_2^4)+
\frac{Z\,\pi}{2}\,[\delta^3(r_1)+\delta^3(r_2)]
-\frac{1}{2}\,p_1^i\,
\biggl(\frac{\delta^{ij}}{r}+\frac{r^i\,r^j}{r^3}\biggr)\,p_2^j \nonumber\\
&&-\frac{Z}{2}\,\frac{m}{M}\,\biggl[
p_1^i\,\biggl(\frac{\delta^{ij}}{r_1} + \frac{r_1^i\,r_1^j}{r_1^3}\biggr)+
p_2^i\,\biggl(\frac{\delta^{ij}}{r_2} + \frac{r_2^i\,r_2^j}{r_2^3}\biggr)\biggr]\,P^j\,,\label{48}
\\
H_B^M & = & \left[
\frac{Z}{4}\biggl(
\frac{\vec{ r}_1}{r_1^3}\times\vec{ p}_1+
\frac{\vec{ r}_2}{r_2^3}\times\vec{ p}_2\biggr)-
\frac{3}{4}\,\frac{\vec{ r}}{r^3}\times(\vec{ p}_1-\vec{ p}_2)
+\frac{Z}{2}\,\frac{m}{M}\,\biggl(
\frac{\vec r_1}{r_1^3} + \frac{\vec r_2}{r_2^3}\biggr)\times\vec{P}
\right]\,\frac{\vec{\sigma}_1+\vec{\sigma}_2}{2}\,,\nonumber\\ \label{49} \\
H_C^M& = & \left[
\frac{Z}{4}\biggl(
\frac{\vec{ r}_1}{r_1^3}\times\vec{ p}_1-
\frac{\vec{ r}_2}{r_2^3}\times\vec{ p}_2\biggr)+
\frac{1}{4}\,\frac{\vec{ r}}{r^3}\times
(\vec{ p}_1+\vec{ p}_2)
+\frac{Z}{2}\,\frac{m}{M}\,\biggl(
\frac{\vec r_1}{r_1^3} - \frac{\vec r_2}{r_2^3}\biggr)\times\vec{P}
\right]\,\frac{\vec{\sigma}_1-\vec{\sigma}_2}{2}\,,\label{50}\nonumber \\ \\
H_D^M  & = & \frac{1}{4}\left(
\frac{\vec{\sigma}_1\,\vec{\sigma}_2}{r^3}
-3\,\frac{\vec{\sigma}_1\cdot\vec{r}\,
\vec{\sigma}_2\cdot\vec{r}}{r^5}\right)\label{51}\,.
\end{eqnarray}
The corresponding second-order correction is
\begin{equation}\label{52}
A_M = \sum_{I=A,B,C,D} \langle H_I^M\,\frac{1}{(E_M-H_M)'}\,H_I^M\rangle_M\,,
\end{equation}
whereas the first-order contribution is given by
\begin{equation}\label{53}
B_M=\langle H_M^{(6)}\rangle_M.
\end{equation}
$H_M^{(6)}$ consists of eleven parts according to Eq. (\ref{24})
with $H_1^M\ldots \,H_7^M$ already defined and
\begin{eqnarray}
H_{8}^M &=&  Z^3\,\frac{m}{M}\,\biggl(4\,\ln2 -\frac{7}{2}\biggr)\,\bigl[\delta^3(r_1)+\delta^3(r_2)\bigr]\label{55}\,,\\
H_{9}^M &=& Z^2\,\frac{m}{M}\,\left(\frac{35}{36} - \frac{448}{27 \pi^2} - 2 \ln(2)
            +\frac{6 \zeta(3)}{\pi^2}\right)\,\bigl[\delta^3(r_1)+\delta^3(r_2)\bigr]\label{56}\,,\\
H_{10}^M &=& \pi\,Z^2\,\biggl(\frac{139}{32}- 2\,\ln(2)+\frac{5}{48} \biggr)\,
              \bigl[\delta^3(r_1)+\delta^3(r_2)\bigr]\,, \label{57}\\
H_{11}^M &=&  \frac{Z}{\pi}\,\biggl(-\frac{4358}{1296} -\frac{10}{27}\, \pi^2 +
              \frac{3}{2}\, \pi^2\,\ln(2) -\frac{9}{4}\, \zeta(3)\biggr)\,
              \bigl[\delta^3(r_1)+\delta^3(r_2)\bigr]\label{58}\,.
\end{eqnarray}
Here $H_8^M$ is the high-energy pure recoil correction taken from hydrogenic results,
$H_9^M$ stands for the radiative recoil correction, and $H_{10}^M$ and $H_{11}^M$ stand for
the one-loop and two-loop radiative corrections, correspondingly \cite{eides}.

% Further calculations are described by the evaluation scheme of Fig. \ref{fig1}
% \begin{figure}
% \includegraphics[scale=0.5]{fig.pdf}
% \caption{Evaluation scheme}
% \label{fig1}
% \end{figure}

\section{Elimination of singularities}
The principal problem of the used approach is that both the first-order and the second-order
contributions are divergent; the divergence cancels out only in the sum of these contributions. To
achieve the cancellation of the divergences, we (i) regularize the divergent contributions by
switching to $d = 3-2\,\epsilon$ dimensions, (ii) move singularities from the second-order
contributions to the first-order ones, and (iii) cancel algebraically the $1/\epsilon$ terms. Moreover,
we notice that the recoil corrections are of two types: (i) corrections due to the perturbation
of the wave function $\phi$, the energy of the reference state $E$, and the nonrelativistic
Hamiltonian $H$ by the nuclear kinetic energy $\vec{P}^2/(2M)$, and (ii) corrections due to the
extra recoil operators in $H^{(4)}_M$ and $H^{(6)}_M$. We will use this fact in the following
derivations.

\subsection{Recoil correction from the second-order contribution}\label{section A}
In this subsection we consider the recoil correction coming from the second-order matrix elements,
i.e. the first term in Eq.~(\ref{08}), which is denoted by $A_M$. The recoil correction from the
second term in Eq.~(\ref{08}), denoted by $B_M$, will be examined in the next subsection.

The second-order contribution with $H_A^M$ is divergent and has to be regularized. Regularization
is performed by rewriting $H_A^M$ in such a way that the singularities are moved from the
second-order matrix element into the first-order ones, where they cancel each other. To do this, we
write $H_A^M$ as
\begin{eqnarray}\label{59}
H_A^M &=& H_{R}^M  - \frac14\,\biggl\{ H_M-E_M,\frac{Z}{r_1}+\frac{Z}{r_2}-\frac{2}{r}
-3\,\frac{m}{M}\biggl(\frac{Z}{r_1}+\frac{Z}{r_2}\biggr)\biggr\}\nonumber\\
    &=& H_R^M + \bigl\{H_M-E_M,\,Q_M\bigr\}\,.
\end{eqnarray}
The operator $Q_M$ is the same as in \cite{pachucki:06hesinglet} with the exception that it also
includes a recoil part $\delta_M Q$. The regular part of operator $H_A^M$ can be evaluated to yield
\begin{equation}\label{60}
H_{R}^M  = H_{R} + \frac{m}{M}\delta_M H_{R}\,,
\end{equation}
\begin{equation}\label{61}
H_{R}\,|\phi\rangle = \biggl\{- \frac{1}{2}(E-V)^2
-\frac{Z}{4}\frac{\vec{r}_1\cdot\vec{\nabla}_1}{r_1^3}
-\frac{Z}{4}\frac{\vec{r}_2\cdot\vec{\nabla}_2}{r_2^3}
+\frac14\,\nabla_1^2\,\nabla_2^2
-p_1^i\,V^{ij}(r)\,p_2^j\biggr\}|\phi\rangle \,,
\end{equation}
\begin{eqnarray}\label{62}
\delta_M H_{R}\,|\phi\rangle &=& \biggl\{(E-V)\biggl(\frac{\vec{P}^2}{2}-\biggl\langle
  \frac{\vec{P}^2}{2} \biggr\rangle\biggr)
+ \frac{3Z}{4}\frac{\vec{r}_1\cdot\vec{\nabla}_2}{r_1^3}
+ \frac{3Z}{4}\frac{\vec{r}_2\cdot\vec{\nabla}_1}{r_2^3}
%+ \frac{\vec{r}\cdot(\vec{\nabla}_1-\vec{\nabla}_2)}{r^3}
 \nonumber \\ &&
- Z\,p_1^i\,V^{ij}(r_1)\,P^j
- Z\,p_2^i\,V^{ij}(r_2)\,P^j\biggr\}|\phi\rangle\,,
\end{eqnarray}
where
\begin{equation}\label{63}
V = -\frac{Z}{r_1}-\frac{Z}{r_2} + \frac{1}{r},
\end{equation}
\begin{equation}\label{64}
V^{ij}(x)=\frac{1}{2x}\biggl(\delta^{ij}+\frac{x^ix^j}{x^2}\biggr)\,.
\end{equation}
Moreover, the kinetic energy of the nucleus is $\langle\vec{P}^2/2\rangle=\delta_M E$. After
regularization, the first term in Eq.~(\ref{08}) takes the form
\begin{eqnarray}\label{65}
A_M &=&
\sum_{a=R,B,C,D}\biggl\langle H_a^M\,\frac{1}{(E_M-H_M)'}\,H_a^M\biggr\rangle_M\nonumber\\
&&+\,\bigl\langle Q_M\,(H_M-E_M)\,Q_M\bigr\rangle_M
+2\,E_M^{(4)}\,\bigl\langle Q_M\bigr\rangle_M
-2\,\bigl\langle H_M^{(4)}\,Q_M\bigr\rangle_M \nonumber\\
&=& A_1^M+A_2^M \,.
\end{eqnarray}
where $A_1^M$ stands for the first term (i.e. the second-order contribution), and $A_2^M$
incorporates the remaining first-order matrix elements. Recoil corrections are obtained by
perturbing the second-order matrix element by the kinetic energy of the nucleus and keeping the
first-order terms in the nuclear mass. So, $\delta_M A_1$ is
\begin{eqnarray}\label{66}
&&\delta_M A_1=
\sum_{a=R,B,C,D}\biggl\langle H_a\frac{1}{(E-H)'}\,\biggl[\frac{ \vec{P}^2}{2}-\delta_M E\biggr]\,\frac{1}{(E-H)'}\,H_a\biggr\rangle\nonumber\\
&&+\,2\,\biggl\langle H_a\,\frac{1}{(E-H)'}\,[\,H_a-\langle H_a\rangle\,]\,\frac{1}{(E-H)'}\,\frac{\vec{P}^2}{2}\biggr\rangle
+\,2\,\biggl\langle \delta_M H_a\frac{1}{(E-H)'}H_a\biggr\rangle,
\end{eqnarray}
while the first-order terms are
\begin{eqnarray}\label{67}
A_2^M&=&\langle Q\,(H_M-E_M)\,Q\rangle_M + 2\,E_M^{(4)}\langle Q\rangle_M - 2\,\langle H^{(4)}_M\,Q\rangle_M\nonumber\\
&&+\,\frac{m}{M}\,\biggl\{2\,\langle Q\,(H-E)\,\delta_M Q\rangle+2E^{(4)}\langle\delta_M Q\rangle-2\,\langle H_A\,\delta_M Q\rangle\biggr\}\,.
\end{eqnarray}
Reduction of these terms will be left to the Appendix, and we present here
the final result for the recoil part
\begin{eqnarray}
\delta_M A_{2}&=&\delta_M\,\biggl\langle-\frac{3}{32}\biggl(\frac{Z^2}{r_1^4}+\frac{Z^2}{r_2^4}\biggr)-\frac{1}{4r^4}
+\frac14\biggl(\frac{Z\vec{r}_1}{r_1^3}-\frac{Z\vec{r}_2}{r_2^3}\biggr)\cdot\frac{\vec{r}}{r^3}+2\,E^{(4)}\,Q\nonumber\\
&&+\,\frac{Z(Z-2)}{4}\pi\,\biggl[\frac{\delta^3(r_1)}{r_2}+\frac{\delta^3(r_2)}{r_1}\biggr]
-\frac14\,p_1^i\biggl(\frac{Z}{r_1}+\frac{Z}{r_2}-\frac{2}{r}\biggr)\frac{1}{r}\biggl(\delta^{ij}+\frac{r^ir^j}{r^2}\biggr)\,p_2^j\nonumber\\
&&+\,\frac12 \,\biggl[p_1^i,\biggl[p_2^j,\frac{1}{r}\biggr]\biggr]\frac{1}{2r}\biggl(\delta^{ij}+\frac{r^ir^j}{r^2}\biggr)
+(E-V)^2\,Q+\frac18\,p_1^2\biggl(\frac{Z}{r_1}+\frac{Z}{r_2}\biggr)\,p_2^2\nonumber\\
&&-\,\frac14\,p_1^2\,\frac{1}{r}\,p_2^2-\frac18[p_1^2,[p_2^2,V]]\biggr\rangle
+\delta_M E^{(4)}\biggl(E+\biggl\langle\frac{1}{2r}\biggr\rangle\biggr)\nonumber\\
&&+\,\biggl\langle\frac{11}{32}\biggl(\frac{Z^2}{r_1^4}+\frac{Z^2}{r_2^4}\biggr)-\frac{3}{16}\frac{Z^2\,\vec{r}_1\cdot\vec{r}_2}{r_1^3r_2^3}
+\frac32\frac{E^{(4)}}{r}-3\,E E^{(4)}+\frac34\,(E-V)^2\biggl(\frac{Z}{r_1}+\frac{Z}{r_2}\biggr)\nonumber\\
&&-\,\frac38\,p_1^2\biggl(\frac{Z}{r_1}+\frac{Z}{r_2}\biggr)\,p_2^2+\frac34\,p_1^i\,\biggl(\frac{Z}{r_1}+\frac{Z}{r_2}\biggr)
\frac{1}{r}\biggl(\delta^{ij}+\frac{r^ir^j}{r^2}\biggr)\,p_2^j+2\,\delta_M E\,(E-V)\,Q\nonumber\\
&&+\,\frac{\pi\,Z}{4}\delta^3(r_1)\,\biggl(\frac{Z-6}{r_2}+2\,E+2\,Z^2\biggr)
+\frac{\pi\,Z}{4}\delta^3(r_2)\,\biggl(\frac{Z-6}{r_1}+2\,E+2\,Z^2\biggr)
\nonumber\\
&&+\,\vec{P}\biggl[\frac{E}{4}\biggl(\frac{Z}{r_1}+\frac{Z}{r_2}\biggr)
-\frac{E}{2r}+\frac14\biggl(\frac{Z}{r_1}+\frac{Z}{r_2}\biggr)^2
-\frac{3}{4r}\biggl(\frac{Z}{r_1}+\frac{Z}{r_2}\biggr)+\frac{1}{2r^2}\biggr]\vec{P}\nonumber\\
&&-\,\sum_a\frac{Z}{4} P^i\left(\frac{\delta^{ij}}{r_a}+\frac{r_a^ir_a^j}{r_a^3}\right)\left(\frac{Z}{r_1}+\frac{Z}{r_2}-\frac{2}{r}\right)p_a^j
\biggr\rangle.\label{77}
\end{eqnarray}

\subsection{Recoil correction from the first-order terms }
In this section we examine the recoil correction coming from the first-order matrix elements, i.e.
the second term in Eq. (\ref{08}), which is denoted as $B_M$. Using Eq. (\ref{24}), $B_M$ can be
written as
\begin{eqnarray}\label{79}
B_M &=& \langle H_M^{(6)}\rangle_M =
\sum_{i=1\ldots11} \langle H_i^M\rangle_M.
\end{eqnarray}
For each of the operators $H_i^M=H_i+\frac{m}{M}\,\delta_M H_i$, the recoil correction is the sum
of two parts: (i) perturbation of the nonrelativistic wave function, $E$ and $H$ by the nuclear
kinetic energy in the non-recoil part, and (ii) the expectation value of the recoil part 
$\delta_M H_i$ (if present). The derivation is straightforward but tedious, so we have moved the description of
this calculation to the Appendix and present only the final result for the recoil correction $\delta_M
B$,
\begin{eqnarray}\label{80}
\delta_M B&=&\delta_M\,\biggl\langle\frac{7}{32}\left(\frac{Z^2}{r_1^4}+\frac{Z^2}{r_2^4}\right)
-\frac{25}{48}\left(\frac{Z\vec{r}_1}{r_1^3}-\frac{Z\vec{r}_2}{r_2^3}\right)\cdot\frac{\vec{r}}{r^3}
+\frac{1}{4}\left(\frac{Z\vec{r}_1}{r_1^3}-\frac{Z\vec{r}_2}{r_2^3}\right)\cdot\frac{\vec{r}}{r^2}-\frac{1}{4r^3}\nonumber\\
&&+\,\frac{41}{48r^4}+\frac{11}{96}\left[p_2^2,\left[p_1^2,\frac{1}{r}\right]\right]+\frac12 (E-V)^3-\frac{3}{8}p_1^2\,(E-V)\,p_2^2\nonumber\\
&&-\,\frac{3}{8}\pi Z\biggl[\,2\left(E+\frac{Z-1}{r_2}\right)\delta^3(r_1)
+2\left(E+\frac{Z-1}{r_1}\right)\delta^3(r_2)-p_1^2\,\delta^3(r_2)-p_2^2\,\delta^3(r_1)\,\biggr]\nonumber\\
&&-\,\frac{\pi}{12}\nabla^2\delta^3(r)
+\frac12\,p_1^i\,\bigl(E-V\bigr)\,\frac{1}{r}\left(\delta^{ij}+\frac{r^ir^j}{r^2}\right)p_2^j
-\frac{1}{8}\frac{Z^2\,r_1^ir_2^j}{r_1^3r_2^3}\left(\frac{r^ir^j}{r}-3\,\delta^{ij}r\right)\nonumber\\
&&-\,\frac{Z}{8}\biggl[\,\frac{r_1^i}{r_1^3}\,p_2^k\left(\delta^{jk}\frac{r^i}{r}
-\delta^{ik}\frac{r^j}{r}-\delta^{ij}\frac{r^k}{r}-\frac{r^ir^jr^k}{r^3}\right)p_2^j
+(1\leftrightarrow 2)\,\biggr]\nonumber\\
&&+\,\frac{1}{8}\,p_1^k\,p_2^l\biggl[-\frac{\delta^{il}\delta^{jk}}{r}+\frac{\delta^{ik}\delta^{jl}}{r}-\frac{\delta^{ij}\delta^{kl}}{r}-\frac{\delta^{jl}r^ir^k}{r^3}
-\frac{\delta^{ik}r^jr^l}{r^3}+3\,\frac{r^ir^jr^kr^l}{r^5}\,\biggr]p_1^i\,p_2^j\nonumber\\
&&+\,\frac{1}{4}\biggl(\vec{p}_1\,\frac{1}{r^2}\,\vec{p}_1+\vec{p}_2\,\frac{1}{r^2}\,\vec{p}_2\biggr)+H_{10}+H_{11}\biggr\rangle\nonumber\\
&&+\,\biggl\langle\frac{3}{2}\,\delta_M E\,(E-V)^2-\frac{3}{4}\,\vec{P}\,(E-V)^2\,\vec{P}-\frac{3}{8}\,\delta_M E\,p_1^2\,p_2^2+\frac{3}{16}\,P^2p_1^2p_2^2\nonumber\\
&&-\,\frac{3}{4}\,\biggl(\delta_M E+3\,E+\frac{3\,(Z-1)}{r_2}-\vec{p}_1\cdot\vec{p}_2\biggr)\pi Z\,\delta^3(r_1)+(1\leftrightarrow2)\nonumber\\
&&+\,\frac{1}{2}\,\delta_M E\,p_1^i\,\frac{1}{r}\left(\delta^{ij}+\frac{r^ir^j}{r^2}\right)p_2^j
-\frac{1}{4}\,\vec{P}^2\,p_1^i\,\frac{1}{r}\left(\delta^{ij}+\frac{r^ir^j}{r^2}\right)\,p_2^j
+\frac{13}{32}\biggl(\frac{Z^2}{r_1^4}+\frac{Z^2}{r_2^4}\biggr)\nonumber\\
&&+\,\frac{13}{16}\,\frac{Z^2\,\vec{r}_1\cdot\vec{r}_2}{r_1^3r_2^3}\biggr\rangle+\langle\delta_M H^{(6)}\rangle\,,
\end{eqnarray}
where
\begin{eqnarray}\label{81}
\langle\delta_M H^{(6)}\rangle&=&
\biggl\langle \frac{Z}{2}\left[p_1^i\,(E-V)\left(\frac{\delta^{ij}}{r_1}+\frac{r_1^ir_1^j}{r_1^3}\right)
+p_2^i\,(E-V)\left(\frac{\delta^{ij}}{r_2}+\frac{r_2^ir_2^j}{r_2^3}\right)\right] P^j
\nonumber\\
&&-\,\frac{Z}{4}\biggl[\,p_1^i\,p_2^k\left(\frac{\delta^{ij}}{r_1}+\frac{r_1^ir_1^j}{r_1^3}\right)p_2^k\, P^j
+p_2^i\,p_1^k\left(\frac{\delta^{ij}}{r_2}+\frac{r_2^ir_2^j}{r_2^3}\right)p_1^k\,P^j\,\biggr]
+\frac{Z^2}{6}\,\frac{\vec{r}_1\cdot\vec{r}_2}{r_1^3r_2^3}\nonumber\\
&&+\,\frac{Z}{4}\biggl[\,p_2^i\left(\frac{\delta^{ij}}{r}+\frac{r^ir^j}{r^3}\right)\left(\frac{\delta^{jk}}{r_1}+\frac{r_1^jr_1^k}{r_1^3}\right)+
p_1^i\left(\frac{\delta^{ij}}{r}+\frac{r^ir^j}{r^3}\right)\left(\frac{\delta^{jk}}{r_2}+\frac{r_2^jr_2^k}{r_2^3}\right)\biggr]P^k\nonumber\\
&&+\,\frac{Z^2}{4}\biggl[\,\vec{p}_1\,\frac{1}{r_1^2}\,\vec{p}_1
+\vec{p}_2\,\frac{1}{r_2^2}\,\vec{p}_2
+p_1^i\left(\frac{\delta^{ij}}{r_1}+\frac{r_1^ir_1^j}{r_1^3}\right)\left(\frac{\delta^{jk}}{r_2}+\frac{r_2^jr_2^k}{r_2^3}\right)p_2^k\biggr]\nonumber\\
&&+\,\frac{Z^3\,\vec{r}_1\cdot\vec{r}_2}{4r_1^3r_2^2}+\frac{Z^3\,\vec{r}_1\cdot\vec{r}_2}{4r_1^2r_2^3}
+\frac{Z^2}{8}\left(\frac{r_1^i}{r_1^3}+\frac{r_2^i}{r_2^3}\right)\biggl(\frac{r_1^ir_1^j-3\,\delta^{ij}\,r_1^2}{r_1}
-\frac{r_2^ir_2^j-3\,\delta^{ij}\,r_2^2}{r_2}\biggr)\frac{r^j}{r^3}\nonumber\\
&&+\,\frac{Z^2}{8}\biggl[\,p_2^k\,\frac{r_1^i}{r_1^3}\left(-\delta^{ik}\frac{r_2^j}{r_2}+\delta^{jk}\frac{r_2^i}{r_2}-\delta^{ij}\frac{r_2^k}{r_2}-\frac{r_2^ir_2^jr_2^k}{r_2^3}\right)p_2^j
+(1\leftrightarrow 2)\,\biggr]\nonumber\\
&&+\,\frac{Z^3}{4r_1^3}+\frac{Z^3}{4r_2^3}-\frac{Z^2}{8r_1^4}-\frac{Z^2}{8r_2^4}
-\frac{3\,Z^3}{2}[\pi\,\delta^3(r_1)+\pi\,\delta^3(r_2)]
+\delta_M H_8+ \delta_M H_9\biggr\rangle\,.
\end{eqnarray}
At this point we have obtained all the terms contributing to the recoil correction.

\subsection{Cancellation of singularities}
The first-order terms $\delta_M A_2$ and $\delta_M B$ could be further transformed using various identities,
namely
\begin{eqnarray}\label{82}
\left[p_2^2,\left[p_1^2,\frac{1}{r}\right]\right]
 &=&\left(\frac{Z\vec{r}_1}{r_1^3}-\frac{Z\vec{r}_2}{r_2^3}\right)\cdot\frac{\vec{r}}{r^3}-\frac{2}{r^4}+P^iP^j\,\frac{3r^ir^j-\delta^{ij}r^2}{r^5}\,,\\
\frac{1}{r^4}
&=&\frac{1}{r^3}+\frac{1}{2}\left(\vec{p}_1\,\frac{1}{r^2}\,\vec{p}_1+\vec{p}_2\,\frac{1}{r^2}\,\vec{p}_2\right)-\left(E+\frac{Z}{r_1}+\frac{Z}{r_2}\right)\frac{1}{r^2}\nonumber\\
&&-\,\frac{m}{M}\,\biggl(\delta_M E-\frac{\vec{P}^2}{2}\biggr)\frac{1}{r^2}\,,\label{83}\\
\frac{Z^2}{r_1^4}\label{84}
&=&\vec{p}_1\,\frac{Z^2}{r_1^2}\,\vec{p}_1-2\left(E+\frac{Z}{r_1}+\frac{Z}{r_2}-\frac{1}{r}\right)\frac{Z^2}{r_1^2}+p_2^2\,\frac{Z^2}{r_1^2}\nonumber\\
&&-\,2\,\frac{m}{M}\,\biggl(\delta_M E-\frac{\vec{P}^2}{2}\biggr)\frac{Z^2}{r_1^2}\,,\\
p_1^i\left(\frac{\delta^{ij}}{r}+\frac{r^ir^j}{r^3}\right)p_2^j&=&-\,2\,H^{(4)}_M-(E-V)^2+\frac{1}{2}\,p_1^2\,p_2^2
+Z\pi\bigl[\delta^3(r_1)+\delta^3(r_2)\bigr]\nonumber\\
&&-\,2\,\frac{m}{M}\,\biggl[\bigl(E-V\bigr)\,\biggl(\delta_M E-\frac{\vec{P}^2}{2}\biggr)-\delta_M H^{(4)}\biggr]\,,\label{85}\\
\nabla^2\,\delta^3(r)&=&\,2\,\vec{p}\,\delta^3(r)\,\vec{p}\,.\label{86}
%p_1^2\,\frac{1}{r}\,p_2^2 &=& \bigl(E-V\bigr)^2\frac{1}{r}-\vec{P}\cdot\vec{p}\,\frac{1}{r}\,\vec{P}\cdot\vec{p}.
\end{eqnarray}
Using these identities we remove all the remaining singularities and
transform the results into a form suitable for numerical calculation.
The final result for recoil correction is presented in the next section.

\section{Final formula}
The final results are split into seven parts: (i) the second-order and third-order matrix elements
containing $H_R$, (ii) the second-order and third-order matrix elements containing $H_B$, (iii) the
second-order and third-order matrix elements containing $H_C$, (iv) the third-order matrix elements
containing $H_D$, (v) the first-order matrix elements with the reference state and the perturbed
wave function, and (vi) the remaining first-order terms with the exception of (vii) pure recoil, the
radiative recoil and the recoil corrections to one-loop and two-loops radiative corrections.

The final formula is then
\begin{equation}
E_\textrm{recoil} = E_\textrm{i} + E_\textrm{ii} + E_\textrm{iii} + E_\textrm{iv} + E_\textrm{v} + E_\textrm{vi} 
+ E_\textrm{vii}\,,
\end{equation}
\begin{eqnarray}\label{Ei}
E_\textrm{i} &=&  \left\langle H_R\,\frac{1}{(E-H)'}\,\biggl(\frac{ \vec{P}^2}{2}-\delta_M E\biggr)\,\frac{1}{(E-H)'}\,H_R\right\rangle \\
&&+\,2\left\langle H_R\,\frac{1}{(E-H)'}\,[H_R-\langle H_R\rangle]\,\frac{1}{(E-H)'}\,\frac{ \vec{P}^2}{2}\right\rangle
+2\left\langle \delta_M H_R\frac{1}{(E-H)'}H_R\right\rangle,\nonumber\\
E_\textrm{ii} &=&  \left\langle H_B\,\frac{1}{(E-H)}\,\biggl(\frac{ \vec{P}^2}{2}-\delta_M E\biggr)\,\frac{1}{(E-H)}\,H_B\right\rangle\nonumber \\
&&+\,2\left\langle H_B\,\frac{1}{(E-H)}\,H_B\frac{1}{(E-H)}\,\frac{ \vec{P}^2}{2}\right\rangle
+2\left\langle H_B\,\frac{1}{(E-H)}\,\delta_M H_B\right\rangle,\label{Eii}\\
E_\textrm{iii} &=&  \left\langle H_C\,\frac{1}{(E-H)}\,\biggl(\frac{ \vec{P}^2}{2}-\delta_M E\biggr)\,\frac{1}{(E-H)}\,H_C\right\rangle\nonumber \\
&&+\,2\left\langle H_C\,\frac{1}{(E-H)}\,H_C\frac{1}{(E-H)}\,\frac{ \vec{P}^2}{2}\right\rangle
+2\left\langle H_C\,\frac{1}{(E-H)}\,\delta_M H_C\right\rangle,\label{Eiii}\\
E_\textrm{iv} &=&  \left\langle H_D\,\frac{1}{(E-H)}\,\biggl(\frac{ \vec{P}^2}{2}-\delta_M E\biggr)\,\frac{1}{(E-H)}\,H_D\right\rangle\nonumber \\
&&+\,2\left\langle H_D\,\frac{1}{(E-H)}\,H_D\frac{1}{(E-H)}\,\frac{ \vec{P}^2}{2}\right\rangle,\label{Eiv}
\end{eqnarray}
Here
\begin{equation}\label{91}
\delta_M E=\biggl\langle\frac{\vec{P}^2}{2}\biggr\rangle = -E+\langle\vec{p}_1\cdot\vec{p}_2\rangle\,.
\end{equation}
For $E_\textrm{v}$ and $E_\textrm{vi}$ the results can be brought into a more suitable form by introducing set of operators $Q_i$, see Tables \ref{oprsQ} and \ref{oprsQ2},
\begin{eqnarray}\label{Ev}
E_\textrm{v}&=&
-\,\frac{E}{8}\,Z\,\delta_M \langle Q_1\rangle +\frac{1}{8}\,Z\,(1-2\,Z)\,\delta_M \langle Q_3\rangle
+\frac{3}{16}\,Z\,\delta_M \langle Q_4\rangle -\frac{1}{24}\,\delta_M \langle Q_6\rangle\nonumber\\
&&+\,\frac{E^2+2\,E^{(4)}}{4}\,\delta_M \langle Q_7\rangle -\frac{9}{8}\,E\,\delta_M \langle Q_8\rangle
+\frac{7}{8}\,\delta_M \langle Q_9\rangle +\frac{E}{2}\,Z^2\,\delta_M \langle Q_{11}\rangle\nonumber\\
&&+\,E\,Z^2\,\delta_M \langle Q_{12}\rangle  -  E\,Z\,\delta_M \langle Q_{13}\rangle -  Z^2\,\delta_M \langle Q_{14}\rangle
+  Z^3\,\delta_M \langle Q_{15}\rangle - \frac{Z^2}{2}\,\delta_M \langle Q_{16}\rangle\nonumber\\
&&-\,\frac{7}{4}\,Z\,\delta_M \langle Q_{17}\rangle - \frac{9}{16}\,Z\,\delta_M \langle Q_{18}\rangle
+ \frac{Z}{2}\,\delta_M \langle Q_{19}\rangle - \frac{Z^2}{8}\,\delta_M \langle Q_{20}\rangle
+ \frac{Z^2}{4}\,\delta_M \langle Q_{21}\rangle \nonumber\\
&& +\,\frac{Z^2}{4}\,\delta_M \langle Q_{22}\rangle + \frac{13}{8}\,\delta_M \langle Q_{23}\rangle
   + \frac{Z}{2}\,\delta_M \langle Q_{24}\rangle - \frac{1}{96}\,\delta_M \langle Q_{25}\rangle
   - \frac{Z}{4}\,\delta_M \langle Q_{26}\rangle \nonumber\\
&&-\,\frac{E}{8}\,\delta_M \langle Q_{27} \rangle
- \frac{Z}{2}\,\delta_M \langle Q_{28} \rangle +\frac{1}{4}\,\delta_M \langle Q_{29} \rangle
+ \frac{1}{8}\,\delta_M \langle Q_{30}\rangle
\end{eqnarray}
and
\begin{eqnarray}\label{Evi}
E_\textrm{vi}&=&
\biggl\langle-\,\frac32\,E^3 -3\,E E^{(4)} - 2\,E^2\,\delta_M E-\frac{3\,E+\delta_M E+4\,Z^2}{8}\,Z\,Q_1
-\frac{Z\,(8\,Z-3)}{8}\,Q_3\nonumber\\
&&+\,\frac{3\,E^2+2\,E\,\delta_M E+6\,E^{(4)}+2\,\delta_M E^{(4)}}{4}\,Q_7-\frac98\,\delta_M E\,Q_8
+\frac{2\,E+\delta_M E}{2}\,Z^2\,Q_{11}\nonumber\\
&&+\,(3\,E+\delta_M E)\,(Z^2\,Q_{12}-Z\,Q_{13})
-3\,Z^2\,Q_{14}+\frac52\,Z^3\,Q_{15}-Z^2\,Q_{16}+\frac32\,Z\,Q_{17}\nonumber\\
&&+\,Z^2\,Q_{21}+\frac32\,Z^2\,Q_{22}+\frac32\,Z\,Q_{24}
-\frac{1}{8}\,\delta_M E\,Q_{27}-\frac34\,Z\,Q_{28}+\frac38\,Z\, Q_{31}+\frac{19}{24}\,Z^2 \,Q_{32}\nonumber\\
&&-\,\frac32\,E\,Z\,Q_{34}+\frac12\,E\,Q_{35}
-\frac34\,Z^2\,Q_{36}-Z^2\,Q_{37}+\frac32\,Z\,Q_{38}+\frac{5}{16}\,Q_{39}+\frac{3}{16}\,Q_{40}\nonumber\\
&&-\,\frac14\,Q_{41}+\frac{Z^2}{2}\,Q_{42}+\frac{Z^2}{2}\,Q_{43}  - \frac{Z}{2}\,Q_{44}
+\frac{Z}{2}\,Q_{45} + \frac{Z^2}{4}\,Q_{46} + \frac{Z^3}{2}\,Q_{47} + \frac{Z^2}{4}\,Q_{48} \nonumber\\
&&-\,\frac{Z^2}{4}\,Q_{49} + \frac{Z^2}{4}\,Q_{50}\biggr\rangle\,.
\end{eqnarray}
Finally,
\begin{equation}\label{Evii}
E_\textrm{vii}=\langle \,\delta_M H_8+ \delta_M H_9\,\rangle
+\delta_M\,\langle \,H_{10}+H_{11}\,\rangle.
\end{equation}

\section{Numerical calculations of matrix elements}

The helium wave function for triplet states is expanded in a basis set of exponential functions in the form of \cite{Korobov}
\begin{eqnarray}\label{95}
\phi(^3S)&=&\sum_{i=1}^{\mathcal{N}} v_i\bigl[e^{-\alpha_i r_1-\beta_i r_2-\gamma_i r}-(r_1\leftrightarrow r_2)\bigr]\,,\\
\phi(^3P)&=&\sum_{i=1}^{\mathcal{N}} v_i\bigl[\vec r_1\,e^{-\alpha_i r_1-\beta_i r_2-\gamma_i r}-(r_1\leftrightarrow r_2)\bigr]\,,
\end{eqnarray}
where $\alpha_i$, $\beta_i$, and $\gamma_i$ are generated quasi-randomly with conditions:
\begin{eqnarray}\label{96}
A_1<\alpha_i<A_2,\,\,\,\beta_i+\gamma_i>\varepsilon, \nonumber \\
B_1<\beta_i<B_2,\,\,\,\alpha_i+\gamma_i>\varepsilon, \nonumber \\
C_1<\gamma_i<C_2,\,\,\,\alpha_i+\beta_i>\varepsilon. \label{98}
\end{eqnarray}
In order to obtain a highly accurate representation of the wave function, following Korobov
\cite{Korobov}, we use a double set of the nonlinear parameters of the form (\ref{95}). The
parameters $A_i$, $B_i$, $C_i$, and $\varepsilon$ are determined by the energy minimization, with
the condition that $\varepsilon>0$, which follows from the normalizability of the wave function.
The linear coefficients $v_i$ in Eq. (\ref{95}) form a vector $v$, which is a solution of the
generalized eigenvalue problem
\begin{eqnarray}\label{99}
H\,v = E\,N\,v\,,
\end{eqnarray}
where $H$ is the matrix of the Hamiltonian in this basis, $N$ is the normalization (overlap)
matrix, and $E$ the eigenvalue, the energy of the state corresponding to $v$. For the solution of
the eigenvalue problem with $\mathcal{N}=100,\,300,\,600,\,900,\,1200,\,1500$ we use a Cholesky
decomposition in octuple precision. As a result we obtain the following nonrelativistic
energies in au
\begin{eqnarray}\label{100}
E(2^3S) = &&-2.175\,229\,378\,236\,791\,306\,, \\
E(2^3P) = &&-2.133\,164\,190\,779\,283\,199\,.\label{101}
\end{eqnarray}
The calculation of matrix elements of the nonrelativistic Hamiltonian is based on the single master
integral,
\begin{eqnarray}\label{102}
\frac{1}{16\pi^2}\int d^3r_1\int d^3r_2\frac{e^{-\alpha r_1-\beta r_2-\gamma r}}{r_1 r_2 r}=
\frac{1}{(\alpha+\beta)(\beta+\gamma)(\gamma+\alpha)}\,.
\end{eqnarray}
The integrals with any additional powers of $r_i$ in the numerator can be obtained by
differentiation with respect to the corresponding parameter $\alpha$, $\beta$ or $\gamma$. The
matrix elements of relativistic corrections involve inverse powers of $r_1$, $r_2$, $r$. These can
be obtained by integration with respect to a corresponding parameter, which leads to the following
formulas
\begin{eqnarray}\label{103}
\frac{1}{16\pi^2}\int d^3r_1\int d^3r_2\frac{e^{-\alpha r_1-\beta r_2-\gamma r}}{r_1 r_2 r^2}&=&
\frac{1}{(\beta+\alpha)(\alpha+\beta)}\ln\biggl(\frac{\beta+\gamma}{\alpha+\gamma}\biggr), \\
\frac{1}{16\pi^2}\int d^3r_1\int d^3r_2\frac{e^{-\alpha r_1-\beta r_2-\gamma r}}{r_1^2 r_2 r^2}&=&\frac{1}{2\beta}\biggl[\,\frac{\pi^2}{6}
+\frac{1}{2}\ln^2\biggl(\frac{\alpha+\beta}{\beta+\gamma}\biggr)\nonumber\\
&&+\,\textrm{Li}_2\biggl(1-\frac{\alpha+\gamma}{\alpha+\beta}\biggr)
+\textrm{Li}_2\biggl(1-\frac{\alpha+\gamma}{\beta+\gamma}\biggr)\biggr]\,.\label{104}
\end{eqnarray}
All matrix elements involved in the $\alpha^6\,m^2/M$ correction, see Tables \ref{oprsQ} and
\ref{oprsQ2}, can be expressed in terms of rational, logarithmic, and dilogarithmic functions, as
above. The high quality of the wave function allowed us to obtain accurate values of the matrix
elements of $Q_i$ and $\delta_M Q_i$ operators. The corresponding numerical results are presented
in Tables \ref{oprsQ} and \ref{oprsQ2}.

For the second-order matrix elements, the inversion of the operator $E-H$ is performed in the basis
of even or odd parity with $l=0,1,2$ and $3$. In the case when the operator acting on the reference
state does not change its symmetry ($H_A$; for $2^3P$, also $H_B$ and $H_D$), it is necessary to
subtract the reference state from the implicit sum over states. This is obtained by the
orthogonalization with respect to the eigenstate with the closest-to-zero eigenvalue of $H-E$. This
eigenvalue is not exactly equal to 0 because we use a basis set with different parameters, which
are obtained by minimization of that particular term.

\section{Results and discussion}

In this paper, we derived the complete recoil contribution of order $\alpha^6\,m^2/M$ to the energy
levels of the triplet states of helium. The final result is given by Eqs. (\ref{Ei}) -
(\ref{Evii}). It is a combination of various contributions
 of two types: (i) perturbations of the nonrelativistic
wave function, energy, and Hamiltonian in the non-recoil matrix elements by the nuclear kinetic
energy operator and (ii) expectation values of extra recoil operators. In Tables \ref{oprsQ} and
\ref{oprsQ2} the matrix elements of individual operators entering Eqs. (\ref{Ev}) - (\ref{Evii})
are presented.

Results of our numerical calculation of $E_\textrm{i}\ldots E_\textrm{vii}$ for the $2^3S_1$ and
$2^3P_1$ states are presented in Table \ref{Es}. For the $2^3S_1$ state, the total
$\alpha^6\,m^2/M$ recoil correction is dominated by the Dirac delta-like term coming from the
one-loop radiative correction, see Eq.~(\ref{57}); the result for the ionization energy being
$-29.91$~$\textrm{kHz}$. Contrary to that, for the $2^3P_1$ state, the contributions from
$E_\textrm{i}\ldots E_\textrm{vii}$ are of similar size but of the opposite sign. 
So, the total correction to the ionization energy is only $-1.11$~$\textrm{kHz}$ in
this case. Contributions of individual recoil terms to the $2^3S-2^3P$ transition energy of helium
are presented in Table \ref{TBL3}.

The obtained results can be used to improve the theoretical prediction of the
${}^3\textrm{He}-{}^4\textrm{He}$ isotope shift of the $2^3S-2^3P$ transition. In this case the
total $m^2/M\,\alpha^6$ recoil correction calculated in this work is $-9.4$~kHz. Individual
contributions for the point nucleus are summarized in Table \ref{TBL5}. In order to estimate the
uncertainty due to omitted higher-order $m^2/M\,\alpha^{7+}$ terms, we considered two typical
contributions. One of them is the hydrogenic recoil $m^2/M\,\alpha^{7+}$ contribution (as evaluated
in \cite{yerokhin:15}) scaled by the expectation value of $\delta(r_1)$ operator; whereas the
second is the hydrogenic $m\,\alpha^{7+}$ contribution with the $\delta(r_1)$ operator perturbed by
$\vec{p}_1\cdot\vec{p}_2$. Since both contributions happen to be small and of opposite sign, we
took the largest one and multiplied it by a conservative coefficient of 2.

The updated theoretical result for the  ${}^3\textrm{He}-{}^4\textrm{He}$ isotope shift  allows us
to improve the accuracy of determination of the nuclear charge radii difference \mbox{$\delta
R^2=R^2(^3\textrm{He})-R^2(^4\textrm{He})$}, derived from the $2^3S-2^3P$ transition 
\cite{pachucki:15:heis}, namely $\delta R^2[\textrm{Cancio\,2012}]=1.069(3)\, \textrm{fm}^2$ and
$\delta R^2[\textrm{Shiner\,95}]=1.061(3)\,\textrm{fm}^2$. This reduces slightly the discrepancy
with the result from the $2^1S-2^3S$ transition \cite{pachucki:15:heis}, $\delta
R^2[\textrm{Rooij\,2011}]=1.028(11)\,\textrm{fm}^2$, but does not remove it entirely. In order to
clarify this further one needs to calculate the complete $\alpha^6\,m^2/M$ recoil correction also
for singlet states of helium.

\begin{acknowledgments}
K.P. and V.P. acknowledge support by the National Science Center (Poland) Grant No.
2012/04/A/ST2/00105, and V.A.Y. acknowledges support by the Ministry of Education and Science of
the Russian Federation (program for organizing and carrying out scientific investigations) and by
RFBR (grant No. 16-02-00538).
\end{acknowledgments}

\appendix

\section{Dimensional regularization}

Since the triplet state wave function vanishes at $r_{12}=0$, the electron-electron operators do
not lead to any singularities, and thus can be calculated directly in $d=3$. There are, however,
several terms arising from the electron-nucleus recoil operators, which need to be treated within
the dimensional regularization in order to isolate the singular part of the operator. We
essentially repeat the approach from \cite{pachucki:06hesinglet}, so only a brief introduction to dimensional
regularization is presented here. The dimension of space is assumed to be $d=3-2\,\epsilon$. The
surface area of the $d$-dimensional unit sphere is
\begin{eqnarray}\label{A1}
\Omega_d=\frac{2\,\pi^{d/2}}{\Gamma(d/2)},
\end{eqnarray}
and the $d$-dimensional Laplacian is
\begin{eqnarray}\label{A2}
\nabla^2=r^{1-d}\partial_r \,r^{d-1}\partial_r.
\end{eqnarray}
The photon propagator, and thus Coulomb interaction, preserves its form in the momentum representation, while in the coordinate representation it is
\begin{eqnarray}
\label{A3}
\mathcal{V}(r) = \int\frac{d^dk}{(2\pi)^d}\,\frac{4\pi}{k^2}\,e^{i\vec{k}\cdot\vec{r}} = \pi^{\epsilon-1/2}\,\Gamma(1/2-\epsilon)\, r^{2\epsilon-1} \equiv \frac{C_1}{r^{1-2\epsilon}}.
\end{eqnarray}
The elimination of singularities will be performed in atomic units. In accordance with \cite{pachucki:06hesinglet}
this is achieved by transformation
\begin{eqnarray}\label{A4}
\vec{r}\rightarrow(m\alpha)^{-1/(1+2\epsilon)}\,\vec{r}
\end{eqnarray}
and pulling factors $m^{(1-2\epsilon)/(1+2\epsilon)}\,\alpha^{2/(1+2\epsilon)}$ and $m^{(1-10\epsilon)/(1+2\epsilon)}\,\alpha^{6/(1+2\epsilon)}$ from $H$ and $H^{(6)}$.
The nonrelativistic Hamiltonian of hydrogen-like systems is
\begin{eqnarray}\label{A5}
H=\frac{\vec{p}\,^2}{2}-Z\frac{C_1}{r^{1-2\epsilon}}\,,
\end{eqnarray}
and that of helium-like systems is
\begin{eqnarray}\label{A6}
H=\frac{\vec{p_1}^2}{2}+\frac{\vec{p_2}^2}{2}-Z\frac{C_1}{r_1^{1-2\epsilon}}-Z\frac{C_1}{r_2^{1-2\epsilon}}+\frac{C_1}{r_{12}^{1-2\epsilon}}\,.
\end{eqnarray}
The solution of the stationary Schr\"{o}dinger equation $H\,\phi=E\,\phi$ is denoted by $\phi$; we
will never need its explicit (and unknown) form in $d$-dimensions. Instead, we will use only the
generalized cusp condition to eliminate various singularities from matrix elements with
relativistic operators. Namely, we expect that for small $r\equiv r_1$
\begin{equation}\label{A7}
\phi(r)\approx\phi(0)\,(1-C\,r^\gamma)
\end{equation}
with some coefficient $C$ and $\gamma$ to be obtained from the two-electron Schr\"{o}dinger equation around $r=0$,
\begin{eqnarray}\label{A8}
\left[\,-\,\frac{\nabla^2}{2}-Z\,\mathcal{V}(r)\,\right]\phi(0)\,(1-C\,r^\gamma)\approx E\,\phi(0)\,(1-C\,r^\gamma).
\end{eqnarray}
From cancellation of small $r$ singularities on the left side of the above equation, one obtains
\begin{eqnarray}\label{A9}
\gamma &=& 1+2\,\epsilon, \\\label{A10}
C &=& -\,\frac{1}{2}\,Z\,\pi^{\epsilon-1/2}\,\Gamma(-1/2-\epsilon).
\end{eqnarray}
Therefore, the two-electron wave function around $r_1=0$ behaves as
\begin{eqnarray}\label{A11}
\phi(\vec{r_1},\vec{r_2})\approx\phi(r_1=0)\,(1-C\,r_1^{1+2\epsilon}).
\end{eqnarray}
Apart from the Coulomb potential $\mathcal{V}(r)$ in the coordinate space, we need also other functions, which appear in the calculations of relativistic operators, namely
\begin{eqnarray}\label{A12}
\mathcal{V}_2(r)&=&\int\frac{d^d k}{(2\pi)^d}\,\frac{4\pi}{k^4}\,e^{i\vec{k}\cdot\vec{r}},\\
\mathcal{V}_3(r)&=&\int\frac{d^d k}{(2\pi)^d}\,\frac{4\pi}{k^6}\,e^{i\vec{k}\cdot\vec{r}}.\label{A13}
\end{eqnarray}
They can be obtained from the differential equations
\begin{eqnarray}
-\nabla^2\mathcal{V}_2(r)&=&\mathcal{V}(r),\label{A14}\\
-\nabla^2\mathcal{V}_3(r)&=&\mathcal{V}_2(r),\label{A15}
\end{eqnarray}
with the result
\begin{eqnarray}\label{A16}
\mathcal{V}_2(r)&=&C_2\,r^{1+2\epsilon},\\
\mathcal{V}_3(r)&=&C_3\,r^{3+2\epsilon},\label{A17}
\end{eqnarray}
where
\begin{eqnarray}\label{A18}
C_2&=&\frac14\,\pi^{\epsilon-1/2}\,\Gamma(-1/2-\epsilon),\\\label{A19}
C_3&=&\frac{1}{32}\,\pi^{\epsilon-1/2}\,\Gamma(-3/2-\epsilon).
\end{eqnarray}
Using $\mathcal{V}_i$, we calculate various integrals involving the photon propagator in the Coulomb
gauge, namely
\begin{eqnarray}
&&\int\frac{d^d k}{(2\pi)^d}\,\frac{4\pi}{k^4}\,\left(\delta^{ij}-\frac{k^i\,k^j}{k^2}\right)\,e^{i\vec{k}\cdot\vec{r}}=\delta^{ij}\,\mathcal{V}_2+\partial^i\partial^j\,\mathcal{V}_3\nonumber\\
&=&\pi^{\epsilon-1/2}\,r^{-1+2\epsilon}\left[\frac{3}{16}\,\delta^{ij}\,\Gamma(-1/2-\epsilon)\,r^2+\frac{1}{8}\,\Gamma(1/2-\epsilon)\,r^i\, r^j\right]\nonumber\\
&\equiv&\left[\frac{1}{8r}\,\bigl(r^i\,r^j-3\,\delta^{ij}\,r^2\bigr)\right]_\epsilon= W_\epsilon^{ij},\label{A20}
\end{eqnarray}
and
\begin{eqnarray}
&&\int \frac{d^d k}{(2\pi)^d}\,\frac{4\pi}{k^2}\,\left(\delta^{ij}-\frac{k^i\,k^j}{k^2}\right)\,e^{i\vec{k}\cdot\vec{r}}=\delta^{ij}\,\mathcal{V}+\partial^i\partial^j\,\mathcal{V}_2\nonumber\\
&=&\pi^{\epsilon-1/2}\,r^{-3+2\epsilon}\left[\frac{1}{2}\delta^{ij}\,\Gamma(1/2-\epsilon)\,r^2+\Gamma(3/2-\epsilon)\,r^i\,r^j\right]\nonumber\\
&\equiv&\left[\frac{1}{2r^3}\,\bigl(\delta^{ij}\,r^2 +r^i\,r^j\bigr)\right]_\epsilon .\label{A21}
\end{eqnarray}
Now we are ready to remove the singularities from matrix elements of various operators. By
convention we pull out a common factor $\bigl[(4\pi)^\epsilon\,\Gamma(1+\epsilon)\bigr]^2$ from all
matrix elements. Then, for example, the matrix element $\langle [Z^3/r^3]_\epsilon\rangle$ with
$r=r_1$ is
\begin{eqnarray}
\left\langle\left[\frac{Z}{r}\right]^3_\epsilon\right\rangle &=& Z^3\,C_1^3\int d^d r\,\phi^2(r)\,r^{-3+6\epsilon} \nonumber\\
&=& Z^3\,C_1^3\,\phi^2(0)\int^a d^d r\,r^{-3+6\epsilon} + Z^3 \int_a d^3 r\,\phi^2(r)\,r^{-3}\nonumber\\
&=&\left\langle\frac{Z^3}{r^3}\right\rangle + Z^3\bigl\langle\pi\,\delta^d (r)\bigr\rangle\left(\frac{1}{\epsilon}+2\right),\label{A22}
\end{eqnarray}
where
\begin{eqnarray}\label{A23}
\left\langle\frac{1}{r^3}\right\rangle
= \lim_{a\rightarrow 0}\int d^3 r\,\phi^2(r)\left[\frac{1}{r^3}\,\Theta(r-a) + 4\pi\,\delta^3(r)\,(\gamma+\ln a)\right]
\end{eqnarray}
is the regularized form of $1/r^3$. The matrix element $\langle [Z^2/r^4]_\epsilon\rangle$ is
\begin{eqnarray}\label{A24}
\left\langle\left[\frac{Z^2}{r^4}\right]_\epsilon\right\rangle&=& Z^2\,C_1^2\int d^d r\,\phi^2(r)\bigl[\nabla\bigl(r^{-1+2\epsilon}\bigr)\bigr]^2\nonumber\\
&=&Z^2\, C_1^2\,(-1+2\epsilon)^2\,\phi^2(0)\int^a d^d r\,r^{-4+4\epsilon}\,\bigl(1-C\,r^{1+2\epsilon}\bigr)^2 + Z^2\int_a d^3r\,\phi^2(r)\,r^{-4}\nonumber \\
&=& \left\langle\frac{Z^2}{r^4}\right\rangle + Z^3\,\bigl\langle\pi\,\delta^d (r)\bigr\rangle\left(-\,\frac{2}{\epsilon}+8\right),
\end{eqnarray}
$\langle 1/r^4\rangle$ in the above is again a regularized form of $1/r^4$,
where $1/a$ and $\ln a+\gamma$ are dropped, analogous to $\langle 1/r^3\rangle$ term. However,
we do not need its explicit form because we can always rewrite it in terms of $\langle Z^3/r^3\rangle$
using expectation value identities. Similarly,
\begin{eqnarray}\label{A25}
Z^2\left\langle\left[\frac{1}{2r}\left(\delta^{ij}+\frac{r^i\,r^j}{r^2}\right)\right]_\epsilon\nabla^i\nabla^j\left[\frac{1}{r}\right]_\epsilon\right\rangle
&=&\left\langle\left[\frac{Z^2}{r^4}\right]_\epsilon\right\rangle + 2\,Z^3\,\bigl\langle\pi\,\delta^d(r)\bigr\rangle\,,
\end{eqnarray}
\begin{eqnarray}\label{A26}
-Z^3\left\langle W^{ij}_\epsilon\,\nabla^i\left[\frac{1}{r}\right]_\epsilon\nabla^j\left[\frac{1}{r}\right]_\epsilon\right\rangle
&=&\frac{1}{4}\left\langle\left[\frac{Z^3}{r^3}\right]_\epsilon\right\rangle - \frac{7\,Z^3}{4}\,\bigl\langle\pi\,\delta^d(r)\bigr\rangle\,,
\end{eqnarray}
and
\begin{eqnarray}
-i\,Z^2\left\langle \nabla^i\left[\frac{1}{r}\right]_\epsilon\,\left[\frac{p^2}{2},W^{ij}_\epsilon\right]\,p^j\right\rangle
&=&\frac{1}{8}\left\langle p^i\,\frac{Z^2}{r^4}\,\bigl(\delta^{ij}\,r^2-3r^ir^j\bigr)\,p^j\right\rangle
\nonumber \\ &&
+ \frac{1}{8}\left\langle\left[\frac{Z^2}{r^4}\right]_\epsilon\right\rangle +
\frac{3\,Z^3}{4}\,\bigl\langle\pi\,\delta^d(r)\bigr\rangle\,.\label{A28}
\end{eqnarray}
The last singular term appearing in these calculations is
\begin{eqnarray}\label{A29}
\left\langle \frac{\sigma^{ij}\sigma^{ij}}{8d}\left[\frac{Z^2}{r^4}\right]_\epsilon\right\rangle = \left\langle \frac{d-1}{8}\left[\frac{Z^2}{r^4}\right]_\epsilon\right\rangle
=\frac{1}{4}\left\langle\left[\frac{Z^2}{r^4}\right]_\epsilon\right\rangle + \frac{Z^3}{2}\,\bigl\langle\pi\,\delta^3(r)\bigl\rangle\,,
\end{eqnarray}
where we used the identity
\begin{equation}
\sigma^{ij}\,\sigma^{ij} = d\,(d-1). \label{A31}
\end{equation}
All the singular terms can now be expressed in terms  $\langle[Z^3/r^3]_\epsilon\rangle$ and $\langle[Z^2/r^4]_\epsilon\rangle$
and using the expectation value identity
\begin{eqnarray}\label{A30}
\left[\frac{Z^2}{r_1^4}\right]_\epsilon=\vec{p}_1\,\frac{Z^2}{r_1^2}\,\vec{p}-2\left(E+\frac{Z}{r_2}-\frac{1}{r}-\frac{p_2^2}{2}\right)\frac{Z^2}{r_1^2}-2\left[\frac{Z^3}{r_1^3}\right]_\epsilon
\end{eqnarray}
they eventually cancel out.

\section{Derivation of $\delta_M A_2$}

Let us present here again the terms contributing to $A_2^M$:
\begin{eqnarray}\label{B1}
A_2^M&=&\langle Q\,(H_M-E_M)\,Q\rangle_M + 2\,E_M^{(4)}\langle Q\rangle_M - 2\,\langle H^{(4)}_M\,Q\rangle_M\nonumber\\
&&+\,\frac{m}{M}\,\biggl\{2\,\langle Q\,(H-E)\,\delta_M Q\rangle+2E^{(4)}\langle\delta_M Q\rangle-2\,\langle H_A\,\delta_M Q\rangle\biggr\}\nonumber\\
&=& A_{2a}^M+A_{2b}^M+A_{2c}^M+A_{2d}^M+A_{2e}^M+A_{2f}^M.
\end{eqnarray}
The first three terms contain both recoil and non-recoil parts, while the latter three are recoil only terms. Individual terms can be
reduced by using expectation value identities:
\begin{eqnarray}\label{B2}
A_{2a}^M&=&\langle Q\,(H_M-E_M)\,Q\rangle_M=\frac{1}{2}\,\langle [Q,[H_M-E_M,Q]]\rangle_M\nonumber\\
&=&\frac12 \,\langle(\nabla_1 Q)^2+(\nabla_2 Q)^2\rangle_M + \frac14 \frac{m}{M}\langle[Q,[\vec{P}^2,Q]]\rangle\nonumber\\
&=&\biggl\langle\frac{1}{32}\biggl(\frac{Z^2}{r_1^4}+\frac{Z^2}{r_2^4}\biggr)+\frac{1}{4r^4}
-\frac18\biggl(\frac{\vec{r}_1}{r_1^3}-\frac{\vec{r}_2}{r_2^3}\biggr)\cdot\frac{\vec{r}}{r^3}\biggr\rangle_M\nonumber\\
&&+\,\frac{m}{M}\biggl\langle\frac{1}{32}\biggl(\frac{Z^2}{r_1^4}+\frac{Z^2}{r_2^4}\biggr)+\frac{1}{16}\frac{Z^2\vec{r}_1\cdot\vec{r}_2}{r_1^3r_2^3}\biggr\rangle,\\
A_{2b}^M&=&2\,E^{(4)}\langle Q\rangle_M+2\,\delta_ME^{(4)}\biggl(\frac{E}{2}+\biggl\langle\frac{1}{4r}\biggr\rangle\biggr),\label{B3}\\
A_{2c}^M&=& -2\,\langle H_M^{(4)}\,Q\rangle_M=X_1+X_2+X_3+X_4,\label{B4}
\end{eqnarray}
where
\begin{eqnarray}
X_4&=&-2\,\langle\delta_M H^{(4)}\,Q\rangle\nonumber\\
&=&\sum_a\biggl\langle -\,\frac{Z}{4}\,P^i\left(\frac{\delta^{ij}}{r_a}+\frac{r_a^ir_a^j}{r_a^3}\right)
\left(\frac{Z}{r_1}+\frac{Z}{r_2}-\frac{2}{r}\right)p_a^j
-\frac{Z^2}{8}\biggl(\frac{\delta^{ij}}{r_a}+\frac{r_a^ir_a^j}{r_a^3}\biggr)\biggl[p_a^i,\biggl[p_a^j,\frac{1}{r_a}\biggr]\biggr]\biggr\rangle\nonumber\\
&=&\sum_a\biggl\langle-\,\frac{Z}{4}\,P^i\biggl(\frac{\delta^{ij}}{r_a}+\frac{r_a^ir_a^j}{r_a^3}\biggr)\biggl(\frac{Z}{r_1}+\frac{Z}{r_2}-\frac{2}{r}\biggr)p_a^j
+\frac14\,\frac{Z^2}{r_a^4} + \frac{Z^3}{2}\,\pi\delta^3(r_a)\biggr\rangle.\label{B5}
\end{eqnarray}
Here we used the identity (\ref{A25}) from Appendix A to rewrite the singular term in
the second equality in (\ref{B5}) as
\begin{eqnarray}
\left\langle\frac{Z^2}{8\,r_a}\left(\delta^{ij}+\frac{r_a^i\,r_a^j}{r_a^2}\right)\nabla_a^i\nabla_a^j\frac{1}{r_a}\right\rangle
&=&\left\langle\frac14\,\frac{Z^2}{r_a^4} + \frac{Z^3}{2}\,\pi\,\delta^3(r_a)\right\rangle\,.
\end{eqnarray}
Further
(using $\langle\,\delta^3(x)/x\,\rangle=0$ which is valid in dimensional regularization)
\begin{eqnarray}
X_{3}&=&-\,\biggl\langle\bigl[Z\,\pi\delta^3(r_1)+Z\,\pi\delta^3(r_2)\bigr]\,Q\biggr\rangle_M
=\biggl\langle\frac{Z\,(Z-2)\,\pi}{4}\biggl(\frac{\delta^3(r_1)}{r_2}+\frac{\delta^3(r_2)}{r_1}\biggr)\biggr\rangle_M,\label{B6}\\
X_{2}&=&\biggl\langle p_1^i\,\frac{1}{r}\biggl(\delta^{ij}+\frac{r^ir^j}{r^2}\biggr)\,p_2^j\,Q\biggr\rangle_M\label{B7}\\
&=&\biggl\langle-\,\frac14\,p_1^i\biggl(\frac{Z}{r_1}+\frac{Z}{r_2}-\frac{2}{r}\biggr)\frac{1}{r}\biggl(\delta^{ij}+\frac{r^ir^j}{r^2}\biggr)\,p_2^j
+\frac12\biggl[p_1^i,\biggl[p_2^j,\frac{1}{r}\biggr]\biggr]\,\frac{1}{2r}\biggl(\delta^{ij}+\frac{r^ir^j}{r^2}\biggr)\biggr\rangle_M,\nonumber\\
X_{1}&=&\frac14\,\bigl\langle\bigl[(p_1^2+p_2^2)-2\,p_1^2p_2^2\bigr]\,Q\bigr\rangle_M\nonumber\\
&=& \frac14\,\bigl\langle(p_1^2+p_2^2)\,Q\,(p_1^2+p_2^2)+\frac12 \,[p_1^2+p_2^2,[Q,p_1^2+p_2^2]]
-2\,p_1^2\,Q\,p_2^2-[p_1^2,[p_2^2,Q]]\bigr\rangle_M\nonumber\\
&=& X_{1A}+X_{1B}+X_{1C}+X_{1D},\label{B8}
\end{eqnarray}
where
\begin{eqnarray}
X_{1A}&=&\langle(E-V)^2\,Q\rangle_M+2\,\frac{m}{M}\,\biggl\langle(E-V)\,Q\,\biggl(\delta_M E-\frac{\vec{P}^2}{2}\biggr)\biggr\rangle\label{B9}\\
&=&\langle(E-V)^2\,Q\rangle_M+\frac{m}{M}\,\biggl\langle2\,\delta_M E\,(E-V)\,Q-\vec{P}\,(E-V)\,Q\vec{P}-\frac12\,[\vec{P},[\vec{P},(E-V)\,Q]]\biggr\rangle\,,\nonumber\\
X_{1B}&=&-\,\frac14\,\biggl\langle\biggl[V+\frac{m}{M}\,\frac{\vec{P}^2}{2},\biggl[p_1^2+p_2^2,\,Q\biggr]\biggr]\biggr\rangle_M\label{B10}\\
&=&\biggl\langle-\,\frac18\,\biggl(\frac{Z^2}{r_1^4}+\frac{Z^2}{r_2^4}\biggr)+\frac38\,\biggl(\frac{Z\vec{r}_1}{r_1^3}-\frac{Z\vec{r}_2}{r_2^3}\biggr)\cdot\frac{\vec{r}}{r^3}
-\frac{1}{2r^4}\biggr\rangle_M\nonumber \\ &&
+\frac{m}{M}\,\biggl\langle\,\frac18\,\biggl(\frac{Z^2}{r_1^4}+\frac{Z^2}{r_2^4}\biggr)+\frac14\,\frac{Z^2\,\vec{r}_1\cdot\vec{r}_2}{r_1^3r_2^3}\biggr\rangle\,,\nonumber\\
X_{1C}&=&\biggl\langle\,\frac18\,p_1^2\,\biggl(\frac{Z}{r_1}+\frac{Z}{r_2}\biggr)\,p_2^2-\frac14\,p_1^2\,\frac{1}{r}\,p_2^2\biggr\rangle_M,\label{B11}\\
X_{1D}&=&\biggl\langle-\,\frac18\,\biggl[p_1^2,\biggl[p_2^2,\frac{1}{r}\biggr]\biggr]\biggr\rangle_M.\label{B12}
\end{eqnarray}
The remaining terms are
\begin{eqnarray}
A_{2d}^M&=&\frac{m}{M}\,\bigl\langle[Q,\,[H-E,\,\delta_M Q]]\bigr\rangle\nonumber\\
&=&\frac{m}{M}\,\bigl\langle(\nabla_1 Q)(\nabla_1\delta_M Q)+(\nabla_2 Q)(\nabla_2\delta_M Q)\bigr\rangle\nonumber\\
&=&\frac{m}{M}\,\biggl\langle-\,\frac{3}{16}\biggl(\frac{Z^2}{r_1^4}+\frac{Z^2}{r_2^4}\biggr)+\frac38\,\biggl(\frac{Z\vec{r}_1}{r_1^3}-\frac{Z\vec{r}_2}{r_2^3}\biggr)
\cdot\frac{\vec{r}}{r^3}\biggr\rangle\,,\label{B13}\\
A_{2e}^M&=&\frac{m}{M}\,\biggl(\,\frac32\,E^{(4)}\,\biggl\langle\frac{1}{r}\biggr\rangle-3\,E E^{(4)}\biggr)\,,\label{B14}\\
A_{2f}^M&=&-2\,\frac{m}{M}\,\langle H_A\,\delta_M Q\rangle\label{B15}
=F_1+F_2+F_3\,,
\end{eqnarray}
where
\begin{eqnarray}
%X_{6D}&=&-\frac{m}{M}\biggl\langle\frac32 \pi\,Z\biggl(\frac{\delta^3(r)}{r_1}+\frac{\delta^3(r)}{r_2}\biggr)\biggr\rangle,\nonumber\\
F_3&=&-\,\frac{m}{M}\,\biggl\langle\frac{3\,Z^2\,\pi}{4}\biggl(\frac{\delta^3(r_1)}{r_2}+\frac{\delta^3(r_2)}{r_1}\biggr)\biggr\rangle,\label{B16}\\
F_2&=&\frac{m}{M}\,\biggl\langle\frac34\, p_1^i\,\biggl(\frac{Z}{r_1}+\frac{Z}{r_2}\biggr)\frac{1}{r}\biggl(\delta^{ij}+\frac{r^ir^j}{r^2}\biggr)\,p_2^j\biggr\rangle,\label{B17}\\
F_1&=&\frac14\,\frac{m}{M}\,\bigl\langle\bigl[(p_1^2+p_2^2)^2-2\,p_1^2p_2^2\bigr]\,\delta Q\bigr\rangle\nonumber\\
&=&\frac14\,\frac{m}{M}\,\bigl\langle(p_1^2+p_2^2)\,\delta Q\,(p_1^2+p_2^2)+\frac12 \,[p_1^2+p_2^2,[p_1^2+p_2^2,\delta Q]]
-2\,p_1^2\,\delta Q\,p_2^2\bigr\rangle\nonumber\\
&=&F_{1A}+F_{1B}+F_{1C}\,,\label{B18}
\end{eqnarray}
and where
\begin{eqnarray}
F_{1A}&=&\frac{m}{M}\,\biggl\langle\frac34\,(E-V)^2\biggl(\frac{Z}{r_1}+\frac{Z}{r_2}\biggr)\biggr\rangle\,,\label{B19}\\
F_{1B}&=&\frac{m}{M}\,\biggl\langle\frac38\,\biggl(\frac{Z^2}{r_1^4}+\frac{Z^2}{r_2^4}\biggr)
-\frac38\,\biggl(\frac{Z\vec{r}_1}{r_1^3}-\frac{Z\vec{r}_2}{r_2^3}\biggr)
\cdot\frac{\vec{r}}{r^3}\biggr\rangle\,,\label{B20}\\
F_{1C}&=&-\,\frac{m}{M}\,\biggl\langle\frac38\, p_1^2\,\biggl(\frac{Z}{r_1}+\frac{Z}{r_2}\biggr)\,p_2^2\biggr\rangle\,.\label{B21}
\end{eqnarray}
Taking now only the recoil part of terms $A_{2a}^M\ldots A_{2f}^M$ we obtain the results:
\begin{eqnarray}\label{68}
\delta_M A_{2a}&=&\delta_M\,\biggl\langle\frac{1}{32}\biggl(\frac{Z^2}{r_1^4}+\frac{Z^2}{r_2^4}\biggr)+\frac{1}{4r^4}
-\frac18\biggl(\frac{\vec{r}_1}{r_1^3}-\frac{\vec{r}_2}{r_2^3}\biggr)\cdot\frac{\vec{r}}{r^3}\biggr\rangle\\
&&+\,\biggl\langle\frac{1}{32}\biggl(\frac{Z^2}{r_1^4}+\frac{Z^2}{r_2^4}\biggr)+\frac{1}{16}\frac{Z^2\,\vec{r}_1\cdot\vec{r}_2}{r_1^3r_2^3}\biggr\rangle,\nonumber\\
\delta_M A_{2b}&=&2\,E^{(4)}\delta_M\,\langle Q\rangle+2\,\delta_ME^{(4)}\biggl(\frac{E}{2}+\biggl\langle\frac{1}{4r}\biggr\rangle\biggr),\label{69}\\
\delta_M A_{2c}&=& \delta_M\,\biggl\langle\frac{Z(Z-2)\,\pi}{4}\biggl(\frac{\delta^3(r_1)}{r_2}+\frac{\delta^3(r_2)}{r_1}\biggr)
-\frac14\,p_1^i\biggl(\frac{Z}{r_1}+\frac{Z}{r_2}-\frac{2}{r}\biggr)\frac{1}{r}\biggl(\delta^{ij}+\frac{r^ir^j}{r^2}\biggr)\,p_2^j\label{70}\nonumber\\
&&+\,\frac12\biggl[p_1^i,\biggl[p_2^j,\frac{1}{r}\biggr]\biggr]\,\frac{1}{2r}\biggl(\delta^{ij}+\frac{r^ir^j}{r^2}\biggr)
+(E-V)^2\,Q -\frac18 \biggl(\frac{Z^2}{r_1^4}+\frac{Z^2}{r_2^4}\biggr)\nonumber\\
&&+\,\frac38 \biggl(\frac{Z\vec{r}_1}{r_1^3}-\frac{Z\vec{r}_2}{r_2^3}\biggr)\cdot\frac{\vec{r}}{r^3}-\frac{1}{2r^4}
+\frac18\,p_1^2\,\biggl(\frac{Z}{r_1}+\frac{Z}{r_2}\biggr)\,p_2^2-\frac14\,p_1^2\,\frac{1}{r}\,p_2^2
-\frac18\biggl[p_1^2,\biggl[p_2^2,\frac{1}{r}\biggr]\biggr]\biggr\rangle\nonumber\\
&&+\,\biggl\langle -\,\frac{Z}{4} \sum_aP^i\left(\frac{\delta^{ij}}{r_a}+\frac{r_a^ir_a^j}{r_a^3}\right)
\biggl(\frac{Z}{r_1}+\frac{Z}{r_2}-\frac{2}{r}\biggr)\,p_a^j
+\frac{3}{8}\biggl(\frac{Z^2}{r_1^4}+\frac{Z^2}{r_2^4}\biggr)\nonumber\\
&& + \,\frac{Z^3}{2}\bigl(\pi\delta^3(r_1)+\pi\delta^3(r_2)\bigr)
+2\,\delta_M E\,(E-V)\,Q-\vec{P}\,(E-V)\,Q\vec{P}\nonumber\\
&&-\,\frac12 \,[\vec{P},[\vec{P},(E-V)\,Q]]+\frac14\frac{Z^2\,\vec{r}_1\cdot\vec{r}_2}{r_1^3r_2^3}\biggr\rangle\,,\nonumber\\
\delta_M A_{2d}&=&\biggl\langle-\,\frac{3}{16}\biggl(\frac{Z^2}{r_1^4}+\frac{Z^2}{r_2^4}\biggr)
+\frac38\biggl(\frac{Z\vec{r}_1}{r_1^3}-\frac{Z\vec{r}_2}{r_2^3}\biggr)
\cdot\frac{\vec{r}}{r^3}\biggr\rangle,\label{74}\\
\delta_M A_{2e}&=&\frac32\,E^{(4)}\biggl\langle\frac{1}{r}\biggr\rangle-3\,E E^{(4)},\label{75}\\
\delta_M A_{2f}&=&\biggl\langle-\,\frac{3\,Z^2\pi}{4}\biggl(\frac{\delta^3(r_1)}{r_2}+\frac{\delta^3(r_2)}{r_1}\biggr)
+\frac34\, p_1^i\,\biggl(\frac{Z}{r_1}+\frac{Z}{r_2}\biggr)\frac{1}{r}\biggl(\delta^{ij}+\frac{r^ir^j}{r^2}\biggr)\,p_2^j\nonumber\\
&&+\,\frac34\,(E-V)^2\biggl(\frac{Z}{r_1}+\frac{Z}{r_2}\biggr)
+\frac38\biggl(\frac{Z^2}{r_1^4}+\frac{Z^2}{r_2^4}\biggr)-\frac38\biggl(\frac{Z\vec{r}_1}{r_1^3}-\frac{Z\vec{r}_2}{r_2^3}\biggr)
\cdot\frac{\vec{r}}{r^3}\nonumber\\
&&-\,\frac38\, p_1^2\,\biggl(\frac{Z}{r_1}+\frac{Z}{r_2}\biggr)\,p_2^2\biggr\rangle\,. \label{76}
\end{eqnarray}
Summing all of the recoil parts $\delta_M A_{2a}\ldots \delta_M A_{2f}$ and using the identity
\begin{eqnarray}\label{78}
[\vec{P},[\vec{P},(E-V)\,Q]]&=&\frac12\,\biggl(\frac{Z^2}{r_1^4}+\frac{Z^2}{r_2^4}\biggr)+\frac{Z^2\,\vec{r}_1\cdot\vec{r}_2}{r_1^3r_2^3}
-\biggl(E+\frac{2\,Z-3}{r_2}\biggr)\pi\,Z\,\delta^3(r_1)\nonumber\\
&&-\,\biggl(E+\frac{2\,Z-3}{r_1}\biggr)\pi\,Z\,\delta^3(r_2)
\end{eqnarray}
we get the final result (\ref{77}).

\section{Derivation of $\delta_M B$}
\noindent In the following we perform only derivation of terms $B_1^M\ldots\,B_7^M$ defined as
\begin{equation}
B_i^M =  \langle H_i^M\rangle_M
\end{equation}
and the evaluation of the remaining terms is trivial since they
contain only Dirac delta-like contributions.
The expectation value of the kinetic term
\begin{equation}\label{C1}
H_1^M=\frac{1}{16}\,\bigl(p_1^6+p_2^6\bigr)
\end{equation}
is
\begin{eqnarray}\label{C2}
B_1^M&=&\frac{1}{16}\,\bigl\langle(p_1^2+p_2^2)^3-3\,p_1^2p_2^2\,(p_1^2+p_2^2)\bigr\rangle_M\nonumber\\
%&=&\frac{1}{4}\,\biggl\langle\biggl(E-V+\frac{m}{M}\biggl(\delta_M E-\frac{\vec{P}_I^2}{2}\biggr)\biggr)\,\bigl(p_1^2+p_2^2\bigr)
%\,\biggl(E-V+\frac{m}{M}\biggl(\delta_M E-\frac{\vec{P}_I^2}{2}\biggr)\biggr)\biggr\rangle_M\nonumber\\
%&&-\frac{3}{8}\,\biggl\langle p_1^2\,p_2^2\,\biggl(E-V+\frac{m}{M}\biggl(\delta_M E-\frac{ P_I^2}{2}\biggr)\biggr)\biggr\rangle_M\nonumber\\
&=&\biggl\langle\frac{1}{8}\,\biggl[V+\frac{m}{M}\frac{\vec{P}^2}{2},\biggl[p_1^2+p_2^2,V\biggr]\biggr]
+\frac{1}{2}\,\biggl(E-V+\frac{m}{M}\biggl(\delta_M E-\frac{\vec{P}^2}{2}\biggr)\biggr)^3\nonumber\\
&&-\,\frac{3}{8}\,p_1^2\,p_2^2\,\biggl(E-V+\frac{m}{M}\biggl(\delta_M E-\frac{\vec{P}^2}{2}\biggr)\biggr)\biggr\rangle_M \nonumber\\
&=& \biggl\langle\frac14\,\bigl[(\nabla_1 V)^2+(\nabla_2 V)^2\bigr]+\frac12\,(E-V)^3-\frac38\,p_1^2\,(E-V)\,p_2^2\nonumber\\
&&+\,\frac{3}{16}\,[p_1^2,[p_2^2,V]]\biggr\rangle_M+\frac{m}{M}\biggl\langle\frac32\,(E-V)^2\biggl(\delta_M E-\frac{\vec{P}^2}{2}\biggr)
-\frac38\,p_1^2p_2^2\biggl(\delta_M E-\frac{\vec{P}^2}{2}\biggr)\nonumber\\
&&-\,\frac12\biggl(\frac{Z^2}{r_1^4}+\frac{Z^2}{r_2^4}\biggr)-\frac{Z^2\,\vec{r}_1\cdot\vec{r}_2}{r_1^3r_2^3}\biggr\rangle\,.
\end{eqnarray}
The recoil correction $\delta_M B_1$ is then
\begin{eqnarray}\label{C3}
\delta_M B_1&=&\delta_M\,\biggl\langle\frac{Z^2}{4r_1^4}+\frac{Z^2}{4r_2^4}-\frac12 \biggl(\frac{Z\vec{r}_1}{r_1^3}-\frac{Z\vec{r}_2}{r_2^3}\biggr)\cdot\frac{\vec{r}}{r^3}
+\frac{1}{2r^4}+\frac12\,(E-V)^3\nonumber\\
&&+\,\frac{3}{16}\,\biggl[p_1^2,\biggl[p_2^2,\frac{1}{r}\biggr]\biggr]
-\frac{3}{8}\,p_1^2\,(E-V)\,p_2^2\biggr\rangle\nonumber\\
&&+\,\biggl\langle\frac32\,\delta_M E\,(E-V)^2-\frac34\,\vec{P}\,(E-V)^2\,\vec{P}+\frac14\biggl(\frac{Z^2}{r_1^4}+\frac{Z^2}{r_2^4}\biggr)
+\frac12\frac{Z^2\,\vec{r}_1\cdot\vec{r}_2}{r_1^3r_2^3}\nonumber\\
&&-\,3\,\biggl(E+\frac{Z-1}{r_2}\biggr)\,\pi\,Z\,\delta^3(r_1)+(1\leftrightarrow2)-\frac38\,p_1^2p_2^2\,\biggl(\delta_M E-\frac{\vec{P}^2}{2}\biggr)\biggr\rangle.
\end{eqnarray}
Here we used
\begin{eqnarray}\label{C4}
[\vec{P},[\vec{P},(E-V)^2]]&=&-\,2\,\biggl(\frac{Z^2}{r_1^4}+\frac{Z^2}{r_2^4}\biggr)-4\,\frac{Z^2\,\vec{r}_1\cdot\vec{r}_2}{r_1^3r_2^3}\nonumber\\
&&+\,2\,(E-V)\,\bigl[4\pi\,Z\delta^3(r_1)+4\pi\,Z\delta^3(r_2)\bigr].
\end{eqnarray}

The operator $H^M_2$ is
\begin{equation}\label{C5}
H_2^M=\sum_a\frac{(\nabla_a V)^2}{8}+\frac{5}{128}\,\bigl[p_a^2,\bigl[p_a^2,V\bigr]\bigr]-\frac{3}{64}\,\bigl\{p_a^2,\nabla_a^2 V\bigr\}.
\end{equation}
For the sake of simplicity we split its expectation value into three parts,
\begin{eqnarray}
B_2^M&&=\biggl\langle\,\frac{1}{8}\,\bigl[(\nabla_1 V)^2+(\nabla_2 V)^2\bigr]
+\frac{5}{128}\,\bigl(\bigl[p_1^2,\bigl[p_1^2,V\bigr]\bigr]+\bigl[p_2^2,\bigl[p_2^2,V\bigr]\bigr]\bigr)
-\frac{3}{32}\,\bigl(p_1^2\,\nabla_1^2V+p_2^2\,\nabla_2^2V\bigr)\biggr\rangle_M\nonumber\\
&&=B_{2a}^M+B_{2b}^M+B_{2c}^M.\label{C6}
\end{eqnarray}
Term $B_{2a}^M=\frac18\langle(\nabla_1V)^2+(\nabla_2V)^2\rangle_M$ needs no further reduction.
The remaining terms could be simplified to
\begin{eqnarray}
B_{2b}^M&=&\frac{5}{128}\,\bigl\langle \bigl[p_1^2+p_2^2,\bigl[p_1^2,V\bigr]\bigr]
+\bigl[p_1^2+p_2^2,\bigl[p_2^2,V\bigr]\bigr]-2\,\bigl[p_1^2,\bigl[p_2^2,V\bigr]\bigr]\bigr\rangle_M\nonumber\\\label{C7}
&=&-\,\frac{5}{64}\,\biggl\langle\biggl[V+\frac{m}{M}\frac{\vec{P}^2}{2},\biggl[p_1^2+p_2^2,V\biggr]\biggr]+\bigl[p_1^2,\bigl[p_2^2,V\bigr]\bigr]\biggr\rangle_M,\\\label{C8}
B_{2c}^M&=&-\,\frac{3}{32}\,\bigl\langle \bigl(p_1^2+p_2^2\bigr)\,\nabla_1^2 V+\bigl(p_1^2+p_2^2\bigr)\,\nabla_2^2 V-p_2^2\,\nabla_1^2 V-p_1^2\,\nabla_2^2 V\bigr\rangle_M\\
&=&-\,\frac{3}{8}\,\pi Z\,\biggl\langle 2\,\biggl[E-V+\frac{m}{M}\,\biggl(\delta_M E-\frac{\vec{P}^2}{2}\biggr)\biggr]\bigl(\delta^3(r_1)+\delta^3(r_2)\bigr)
-p_1^2\,\delta^3(r_2)-p_2^2\,\delta^3(r_1)\biggr\rangle_M\nonumber.
\end{eqnarray}
Taking now only the recoil parts of individual terms we get
\begin{eqnarray}\label{C9}
\delta_M B_{2a}&=&\frac{1}{8}\,\delta_M\,\biggl\langle(\nabla_1 V)^2+(\nabla_2 V)^2\biggr\rangle,\\ \label{C10}
\delta_M B_{2b}&=&-\,\frac{5}{32}\,\delta_M\,\biggl\langle(\nabla_1 V)^2+(\nabla_2 V)^2+\frac{1}{2}\bigl[p_1^2,\bigl[p_2^2,V\bigr]\bigr]
\biggr\rangle
+\frac{5}{64}\,\bigl\langle\bigl[V,\bigl[\vec{P}^2,V\bigr]\bigr]\bigr\rangle,\\
\delta_M B_{2c}&=&-\,\frac{3}{8}\,\pi Z\,\delta_M\,\biggl\langle 2\left(E+\frac{Z-1}{r_2}\right)\delta^3(r_1)+2\left(E+\frac{Z-1}{r_1}\right)\delta^3(r_2)
-p_1^2\,\delta^3(r_2)-p_2^2\,\delta^3(r_1)\biggr\rangle\nonumber\\
&&-\,\frac{3}{4}\,\biggl\langle \biggl(\delta_M E-\frac{\vec{P}^2}{2}\biggr)\,\pi Z\,\bigl(\delta^3(r_1)+\delta^3(r_2)\bigr)\biggr\rangle\,.\label{C11}
\end{eqnarray}
Term $\delta_M B_2$ is then the sum of these three terms and takes the form
\begin{eqnarray}
\delta_M B_2&=&\delta_M\,\biggl\langle
-\,\frac{1}{32}\biggl(\frac{Z^2}{r_1^4}+\frac{Z^2}{r_2^4}\biggr)+\frac{1}{16}\left(\frac{Z\vec{r}_1}{r_1^3}-\frac{Z\vec{r}_2}{r_2^3}\right)\cdot\frac{\vec{r}}{r^3}-\frac{1}{16r^4}
-\frac{5}{64}\biggl[p_1^2,\biggl[p_2^2,\frac{1}{r}\biggr]\biggr]\nonumber\\
&&-\,\frac{3}{8}\,\pi Z\biggl[\,2\left(E+\frac{Z-1}{r_2}\right)\delta^3(r_1)+2\left(E+\frac{Z-1}{r_1}\right)\delta^3(r_2)\nonumber\\
&&-\,p_1^2\,\delta^3(r_2)-p_2^2\,\delta^3(r_1)\biggr]\biggr\rangle
+\biggl\langle\frac{5}{32}\biggl(\frac{Z^2}{r_1^4}+\frac{Z^2}{r_2^4}\biggr)+\frac{5}{16}\frac{Z^2\,\vec{r}_1\cdot\vec{r}_2}{r_1^3r_2^3}\nonumber\\
&&-\,\frac{3}{4}\biggl(\delta_M E-E+\frac{1-Z}{r_2}-\vec{p}_1\cdot\vec{p}_2\biggr)\pi Z \delta^3(r_1)+(1\leftrightarrow2)\biggr\rangle\,. \label{C12}
\end{eqnarray}

Operator $H_3^M$ is
\begin{equation}\label{C13}
H_3^M=\frac{1}{64}\,\biggl[-4\pi\,\nabla^2\delta^3(r)+\frac{4}{3}\,p_1^i\biggl(
\frac{2}{3}\,\delta^{ij}4\pi\,\delta^3(r)+\frac{1}{r^5}\,\bigl(3\,r^ir^j-\delta^{ij}r^2\bigr)\biggr)\,p_2^j\,\biggr].
\end{equation}
Here we used the identity valid for triplets
\begin{equation}\label{sigma.sigma}
\sigma_1^{ij}\sigma_2^{ij}=2\,\vec\sigma_1\cdot\vec\sigma_2=2
\end{equation}
to evaluate the spin product in $H_3^M$. Since there is no
singular term, we can use $d=3$ representation and
the scalar product in the evaluation of the spin part.
This will be assumed also in all the other
terms where the spin product appears. In the
case where the term containing the spin product is singular
and one has to use its $d$-dimensional form to
evaluate such term, it will be explicitly stated.
The expectation value of $H_3^M$ is
\begin{eqnarray}\label{C14}
B_3^M&=&\biggl\langle-\frac{1}{16}\,\pi\,\nabla^2\delta^3(r)+\frac{1}{48}\,p_1^i\biggl(\frac{2}{3}\,\delta^{ij}4\pi\,\delta^3(r)
+\frac{1}{r^5}\,\bigl(3\,r^ir^j-\delta^{ij}r^2\bigr)\biggr)\,p_2^j\biggr\rangle_M\nonumber\\
&=&\left\langle-\frac{1}{48}\,p_1^i\frac{1}{r^5}\,\bigl(\delta^{ij}r^2-3\,r^ir^j\bigr)p_2^j-\frac{13}{144}\pi\,\nabla^2\delta^3(r)\right\rangle_M,
\end{eqnarray}
where we used the expectation value identity
\begin{equation}\label{C15}
\langle\vec{p}_1\,\delta^3(r)\,\vec{p}_2\rangle=-\frac{1}{2}\,\langle\nabla^2\delta^3(r)\rangle\,.
\end{equation}
Further, with the help of identity
\begin{eqnarray}\label{C16}
p_1^i\,\frac{1}{r^5}\,\bigl(\delta^{ij}r^2-3\,r^ir^j\bigr)\,p_2^j
&=&-\,\frac14\biggl[p_1^2,\biggl[p_2^2,\frac{1}{r}\biggr]\biggr]-\frac{\pi}{3}\,\nabla^2\delta^3(r)
\end{eqnarray}
we get the resulting recoil correction $\delta_M B_3$
\begin{eqnarray}\label{C17}
\delta_M B_3&=&\delta_M\,\biggl\langle\frac{1}{192}\biggl[p_1^2,\biggl[p_2^2,\frac{1}{r}\biggr]\biggr]
-\frac{\pi}{12}\nabla^2\delta^3(r)\biggr\rangle.
\end{eqnarray}

We split the correction due to operator $H_4^M=H_4+\frac{m}{M}\,\delta_M H_4$ into two parts: the recoil correction
to operator $H_4$, which we denote as $B^M_{4a}$, and the expectation value of the recoil part $\delta_M H_4$, which we
denote as $B_{4b}^M$.
The non-recoil part of the operator $H_4^M$ is (omitting the part with $\delta^3(r)$, which does not contribute for triplet states)
\begin{equation}\label{C18}
H_4=\frac{1}{4}\,\bigl(p_1^2+p_2^2\bigr)\,p_1^i\frac{1}{r}\left(\delta^{ij}+\frac{r^ir^j}{r^2}\right)p_2^j\,.
\end{equation}
The expectation value of this is
\begin{eqnarray}
B_{4a}^M
%&=&\frac{1}{2}\,\biggl\langle\biggl(H_M-V-\frac{m}{M}\frac{ P_I^2}{2}\biggr)\,p_1^i\,\frac{1}{r}\left(\delta^{ij}+\frac{r^ir^j}{r^2}\right)p_2^j\biggr\rangle_M\nonumber\\
&=&\frac{1}{2}\,\biggl\langle\bigl(E-V\bigr)\,p_1^i\,\frac{1}{r}\biggl(\delta^{ij}+\frac{r^ir^j}{r^2}\biggr)\,p_2^j\biggr\rangle_M
+\frac{m}{2M}\,\biggl\langle\biggl(\delta_M E-\frac{\vec{P}^2}{2}\biggr)\,p_1^i\,\frac{1}{r}\left(\delta^{ij}+\frac{r^ir^j}{r^2}\right)p_2^j\biggr\rangle\nonumber\\
&=&\frac{1}{2}\,\biggl\langle p_1^i\,\bigl(E-V\bigr)\,\frac{1}{r}\biggl(\delta^{ij}+\frac{r^ir^j}{r^2}\biggr)\,p_2^j
-\frac{1}{2r}\biggl(\delta^{ij}+\frac{r^ir^j}{r^2}\biggr)\biggl[p_1^i,\biggl[p_2^j,\frac{1}{r}\biggr]\biggr]\biggr\rangle_M\nonumber\\
&&+\,\frac{m}{2M}\biggl\langle\biggl(\delta_M E-\frac{\vec{P}^2}{2}\biggr)\,p_1^i\,\frac{1}{r}\biggl(\delta^{ij}+\frac{r^ir^j}{r^2}\biggr)p_2^j\biggr\rangle\,.\label{C19}
\end{eqnarray}
Recoil correction $\delta_M B_{4a}$ is then
\begin{eqnarray}
\delta_M B_{4a}&=&\delta_M\,\biggl\langle\,\frac12 \,p_1^i\,\bigl(E-V\bigr)\,\frac{1}{r}\biggl(\delta^{ij}+\frac{r^ir^j}{r^2}\biggr)p_2^j\
-\frac{1}{2r^4}\biggr\rangle\nonumber\\
&&+\,\biggl\langle\,\frac12\,\biggl(\delta_M E-\frac{\vec{P}^2}{2}\biggr)\,p_1^i
\,\frac{1}{r}\biggl(\delta^{ij}+\frac{r^ir^j}{r^2}\biggr)\,p_2^j\biggr\rangle.\label{C20}
\end{eqnarray}
The recoil part of $H_4^M$ is
\begin{equation}\label{C21}
\delta_M H_4=\,\frac{Z}{4}\,\biggl(p_1^2\,p_1^i\biggl(\frac{\delta^{ij}}{r_1}+\frac{r_1^ir_1^j}{r_1^3}\biggr)\,P^j
+p_2^2\,p_2^i\biggl(\frac{\delta^{ij}}{r_2}+\frac{r_2^ir_2^j}{r_2^3}\biggr)\,P^j\biggr).
\end{equation}
The expectation value of this operator can then be reduced to
\begin{eqnarray}
\delta_M B_{4b}&=&\,\frac{Z}{4}\,\biggl\langle2\,\bigl(E-V\bigr)\biggl[p_1^i\biggl(\frac{\delta^{ij}}{r_1}+\frac{r_1^ir_1^j}{r_1^3}\biggr)\,P^j
+p_2^i\biggl(\frac{\delta^{ij}}{r_2}+\frac{r_2^ir_2^j}{r_2^3}\biggr)\,P^j\,\biggr]\nonumber\\
&&-\,\biggl[p_2^2\,p_1^i\biggl(\frac{\delta^{ij}}{r_1}+\frac{r_1^ir_1^j}{r_1^3}\biggr)\,P^j
+p_1^2\,p_2^i\biggl(\frac{\delta^{ij}}{r_2}+\frac{r_2^ir_2^j}{r_2^3}\biggr)\,P^j\,\biggr]\biggr\rangle\nonumber\\
&=&\biggl\langle\,\frac{Z}{2}\left[p_1^i\,\bigl(E-V\bigr)\left(\frac{\delta^{ij}}{r_1}+\frac{r_1^ir_1^j}{r_1^3}\right) P^j
+p_2^i\,\bigl(E-V\bigr)\left(\frac{\delta^{ij}}{r_2}+\frac{r_2^ir_2^j}{r_2^3}\right) P^j\right]\nonumber\\
&&-\,\frac{1}{2}\left(\frac{Z^2}{r_1^4}+\frac{Z^2}{r_2^4}\right)-Z^3\left[\pi\delta^3(r_1)+\pi\delta^3(r_2)\right]\nonumber\\
&&-\,\frac{Z}{4}\biggl[\,p_1^i\,p_2^k\,\left(\frac{\delta^{ij}}{r_1}+\frac{r_1^ir_1^j}{r_1^3}\biggr)\,p_2^k\, P^j
+p_2^i\,p_1^k\,\biggl(\frac{\delta^{ij}}{r_2}+\frac{r_2^ir_2^j}{r_2^3}\right)\,p_1^k\, P^j\,\biggr]\biggr\rangle\,.\label{C22}
\end{eqnarray}
When commuting $E-V$ we used equation (\ref{A25}) of Appendix A,
in particular
\begin{equation}
\biggl\langle\frac{Z}{4}\,[p_a^i,\,[P^j,\,E-V]]\,\biggl(\frac{\delta^{ij}}{r_a}+\frac{r_a^ir_a^j}{r_a^3}\biggr)\biggr\rangle
=\biggl\langle-\frac{1}{2}\,\frac{Z^2}{r_a^4}-Z^3\,\pi\,\delta^3(r_a)\biggr\rangle\,.
\end{equation}
Operator $H_5^M$ is
\begin{equation}\label{C23}
H_5^M=-\,\frac{1}{6}\left(\frac{Z\vec{r}_1}{r_1^3}-\frac{Z\vec{r}_2}{r_2^3}\right)\cdot\frac{\vec{r}}{r^3}+\frac{1}{3r^4}
+\frac{1}{48}\left(\left[p_1^2,\left[p_1^2,\frac{1}{r}\right]\right]+\left[p_2^2,\left[p_2^2,\frac{1}{r}\right]\right]\right),
\end{equation}
where the spin product was again resolved using identity (\ref{sigma.sigma}). The expectation value is
\begin{eqnarray}\label{C24}
B_5^M
%&=&\biggl\langle-\,\frac{1}{6}\biggl(\frac{Z\vec{r}_1}{r_1^3}-\frac{Z\vec{r}_2}{r_2^3}\biggr)\cdot\frac{\vec{r}}{r^3}+\frac{1}{3r^4}
%+\frac{1}{48}\biggl(\biggl[p_1^2+p_2^2,\biggl[p_1^2+p_2^2,\frac{1}{r}\biggr]\biggr]
%-2\biggl[p_1^2,\biggl[p_2^2,\frac{1}{r}\biggr]\biggr]\biggr)\biggr\rangle_M\nonumber\\
&=&\biggl\langle-\,\frac{1}{6}\,\biggl(\frac{Z\vec{r}_1}{r_1^3}-\frac{Z\vec{r}_2}{r_2^3}\biggr)\cdot\frac{\vec{r}}{r^3}+\frac{1}{3r^4}
-\frac{1}{24}\,\biggl(\biggl[V,\biggl[p_1^2+p_2^2,\frac{1}{r}\biggr]\biggr]
+\biggl[p_1^2,\biggl[p_2^2,\frac{1}{r}\biggr]\biggr]\biggr)\biggr\rangle_M\,.
\end{eqnarray}
The recoil correction is then
\begin{eqnarray}\label{C25}
\delta_M B_5&=&\delta_M\,\biggl\langle-\,\frac{1}{12}\left(\frac{Z\vec{r}_1}{r_1^3}-\frac{Z\vec{r}_2}{r_2^3}\right)\cdot\frac{\vec{r}}{r^3}+\frac{1}{6r^4}
-\frac{1}{24}\,\biggl[p_1^2,\biggl[p_2^2,\frac{1}{r}\biggr]\biggr]\biggr\rangle.
\end{eqnarray}

The operator $H_6^M$ contains the recoil part $\delta_M H_6$, so we 
again split the calculation into two parts: the recoil correction due
to $H_6$, denoted as $\delta_M B_{6a}$, and the expectation value of
$\delta_M H_6$, which we denote as $\delta_M B_{6b}$.
The non-recoil part of the operator $H_6^M$ is
\begin{equation}\label{C26}
H_6=\frac{1}{8}\,p_1^i\,\frac{1}{r^2}\biggl(\delta^{ij}+3\,\frac{r^ir^j}{r^2}\biggr)\,p_1^j
+\frac{1}{8}\,p_2^i\,\frac{1}{r^2}\biggl(\delta^{ij}+3\,\frac{r^ir^j}{r^2}\biggr)\,p_2^j+\frac{1}{2r^4},
\end{equation}
where we used the identity from Eq. (\ref{A31}).
%\begin{equation}
%\sigma^{ij}_a\sigma^{ij}_a=6\,.
%\end{equation}
The recoil correction due to this operator is simply
\begin{equation}\label{C27}
\delta_M B_{6a}=\delta_M\,\biggl\langle\frac{1}{8}\,p_1^i\,\frac{1}{r^2}\biggl(\delta^{ij}+3\,\frac{r^ir^j}{r^2}\biggr)\,p_1^j
+\frac{1}{8}\,p_2^i\,\frac{1}{r^2}\biggl(\delta^{ij}+3\frac{r^ir^j}{r^2}\biggr)\,p_2^j+\frac{1}{2r^4}\biggr\rangle.
\end{equation}
The recoil part of operator $H_6^M$ contains a singular term with spin product,
which we have to evaluate using dimensional regularization.
In particular, we use Eq. (\ref{A29}) to get
\begin{equation}
\biggl\langle\frac{\sigma^{ij}_a\sigma^{ij}_a}{24}\,\frac{Z^2}{r_a^4}\biggr\rangle
=\biggl\langle\frac14\,\frac{Z^2}{r_a^4}+\frac{Z^3}{2}\,\pi\,\delta^3(r_a)\biggr\rangle\,.
\end{equation}
Using this and (\ref{sigma.sigma}) the expectation value of $\delta_M H_6$ can be evaluated to
\begin{eqnarray}
\delta_M B_{6b}&=&\biggl\langle\frac{Z}{4}\biggl[\,p_2^i\biggl(\frac{\delta^{ij}}{r}
+\frac{r^ir^j}{r^3}\biggr)\biggl(\frac{\delta^{jk}}{r_1}+\frac{r_1^jr_1^k}{r_1^3}\biggr)+
p_1^i\biggl(\frac{\delta^{ij}}{r}+\frac{r^ir^j}{r^3}\biggr)\biggl(\frac{\delta^{jk}}{r_2}+\frac{r_2^jr_2^k}{r_2^3}\biggr)\,\biggr] P^k\nonumber\\
&&+\,\frac{1}{4}\,\biggl(\frac{Z^2}{r_1^4}+\frac{Z^2}{r_2^4}+\frac{2}{3}\frac{Z^2\,\vec{r}_1\cdot\vec{r}_2}{r_1^3r_2^3}\biggr)
+\frac{Z^3}{2}\biggl[\,\pi\delta^3(r_1)+\pi\delta^3(r_2)\,\biggr]\nonumber\\
&&+\,\frac{Z^2}{8}\biggl[\,p_1^i\,\frac{1}{r_1^2}\biggl(\delta^{ij}+3\,\frac{r_1^ir_1^j}{r_1^2}\biggr)\,p_1^j
+p_2^i\,\frac{1}{r_2^2}\biggl(\delta^{ij}+3\,\frac{r_2^ir_2^j}{r_2^2}\biggr)\,p_2^j\nonumber\\
&&
+\,2\,p_1^i\,\biggl(\frac{\delta^{ij}}{r_1}+\frac{r_1^ir_1^j}{r_1^3}\biggr)\biggl(\frac{\delta^{jk}}{r_2}
+\frac{r_2^jr_2^k}{r_2^3}\biggr)\,p_2^k\,\biggr]\biggr\rangle\,.\label{C28}
\end{eqnarray}

Finally, we calculate correction due to the operator $H_7^M=H_{7a}^M+H_{7c}^M+H_{7d}^M$.
We split it correspondingly into three parts, $B_7^M=B_{7a}^M+B_{7c}^M+B_{7d}^M$.
The operator $H_{7a}^M$ reads
\begin{eqnarray}\label{C29}
H_{7a}^M&=&-\frac{1}{8}\,\biggr\{\bigl[p_1^i,V\bigr]\biggl(\frac{r^i r^j}{r}-3\,\delta^{ij}r\biggr)\bigl[V,p_2^j\bigr]+\bigl[p_1^i,V\bigr]
\biggl[\frac{p_2^2}{2},\,\frac{r^ir^j}{r}-3\,\delta^{ij}r\biggr]p_2^j\nonumber\\
&&+\,p_1^i\biggl[\frac{r^ir^j}{r}-3\,\delta^{ij}r,\,\frac{p_1^2}{2}\biggr]\bigl[V,p_2^j\bigr]
+p_1^i\biggl[\frac{p_2^2}{2},\,\biggl[\frac{r^ir^j}{r}-3\,\delta^{ij}r,\,\frac{p_1^2}{2}\biggr]\biggr]p_2^j\,\biggr\}\,.
\end{eqnarray}
%Term $B_{7a}^M$ can then be again written as the sum of three parts,
%\begin{equation}
%B_{7a}^M=Y_1+Y_2+Y_3,
%\end{equation}
%where
%\begin{eqnarray}
%Y_1 &=&\left\langle-\frac{1}{8}\frac{Zr_1^i}{r_1^3}\frac{Zr_2^j}{r_2^3}\left(\frac{r^ir^j}{r}-3\delta^{ij}r\right)
%+\frac{1}{4}\left(\frac{Z\vec{r}_1}{r_1^3}-\frac{Z\vec{r}_2}{r_2^3}\right)\cdot\frac{\vec{r}}{r^2}-\frac{1}{4r^3}\right\rangle_M,\\
%Y_2&=&\left\langle-\frac{1}{8}\left\{\bigl[p_1^i,V\bigr]\left[\frac{p_2^2}{2},\frac{r^ir^j}{r}-3\delta^{ij}r\right]p_2^j
%+p_1^i\left[\frac{r^ir^j}{r}-3\delta^{ij}r,\frac{p_1^2}{2}\right]\bigl[V,p_2^j\bigr]\right\}\right\rangle_M\nonumber\\
%&=&\left\langle-\frac{Z}{8}\biggl[\frac{r_1^i}{r_1^3}\,p_2^k\,\left(\delta^{jk}\frac{r^i}{r}-\delta^{ik}\frac{r^j}{r}-\delta^{ij}\frac{r^k}{r}-\frac{r^ir^jr^k}{r^3}\right)p_2^j
%+(1\leftrightarrow 2)\biggr]\right.\nonumber\\
%&&\left.+\frac{1}{8}\biggl[p_2^j\,\frac{1}{r^4}\left(\delta^{jk}r^2-3r^jr^k\right)p_2^k+(1\leftrightarrow 2)\biggr]+\frac{1}{4r^4}\right\rangle_M,\\
%Y_3&=&\left\langle\frac{1}{8}p_1^k\,p_2^l\left[-\frac{\delta^{il}\delta^{jk}}{r}+\frac{\delta^{ik}\delta^{jl}}{r}-\frac{\delta^{ij}\delta^{kl}}{r}-\frac{\delta^{jl}r^ir^k}{r^3}
%-\frac{\delta^{ik}r^jr^l}{r^3}+3\frac{r^ir^jr^kr^l}{r^5}\right]p_1^i\,p_2^j\right\rangle_M.
%\end{eqnarray}
The recoil correction due to this operator is
\begin{eqnarray}\label{C30}
\delta_M B_{7a}&=&\delta_M\,\biggl\langle-\,\frac{1}{8}\frac{Zr_1^i}{r_1^3}\frac{Zr_2^j}{r_2^3}\left(\frac{r^ir^j}{r}-3\delta^{ij}r\right)
+\frac{1}{4}\left(\frac{Z\vec{r}_1}{r_1^3}-\frac{Z\vec{r}_2}{r_2^3}\right)\cdot\frac{\vec{r}}{r^2}-\frac{1}{4r^3}\\
&&-\,\frac{Z}{8}\biggl[\,\frac{r_1^i}{r_1^3}\,p_2^k\left(\delta^{jk}\frac{r^i}{r}-\delta^{ik}\frac{r^j}{r}-\delta^{ij}\frac{r^k}{r}-\frac{r^ir^jr^k}{r^3}\right)p_2^j
+(1\leftrightarrow 2)\,\biggr]\nonumber\\
&&+\,\frac{1}{8}\biggl[\,p_2^j\,\frac{1}{r^4}\left(\delta^{jk}r^2-3r^jr^k\right)p_2^k+(1\leftrightarrow 2)\,\biggr]+\frac{1}{4r^4}\nonumber\\
&&+\,\frac{1}{8}\,p_1^k\,p_2^l\biggl[-\frac{\delta^{il}\delta^{jk}}{r}+\frac{\delta^{ik}\delta^{jl}}{r}-\frac{\delta^{ij}\delta^{kl}}{r}-\frac{\delta^{jl}r^ir^k}{r^3}
-\frac{\delta^{ik}r^jr^l}{r^3}+3\,\frac{r^ir^jr^kr^l}{r^5}\,\biggr]p_1^i\,p_2^j\biggr\rangle.\nonumber
\end{eqnarray}
The operator $H_{7c}^M$ is
\begin{equation}\label{C31}
H_{7c}^M=\frac{1}{24}\left[p_2^2,\left[p_1^2,\frac{1}{r}\right]\right],
\end{equation}
where we used (\ref{sigma.sigma}) for the spin part.
The corresponding recoil correction is simply
\begin{eqnarray}\label{C32}
\delta_M B_{7c}&=&\delta_M\,\biggl\langle\frac{1}{24}\left[p_2^2,\left[p_1^2,\frac{1}{r}\right]\right]\biggr\rangle\,.
\end{eqnarray}
Finally, the operator $H_{7d}^M$ is
\begin{eqnarray}
H_{7d}^M&=&i\,\frac{Z^2}{8M}\sum_{a,b}\frac{r_a^i}{r_a^3}\biggl[H-E,\,\frac{r_b^ir_b^j-3\,\delta^{ij}r_b^2}{r_b}\,p_b^j\biggr]\nonumber\\
&=&i\,\frac{Z^2}{8M}\sum_{a,b}\frac{r_a^i}{r_a^3}\biggl\{\bigl[V,p_b^j\bigr]\,\frac{r_b^ir_b^j-3\,\delta^{ij}r_b^2}{r_b}
+\biggl[\frac{p_b^2}{2},\,\frac{r_b^ir_b^j-3\,\delta^{ij}r_b^2}{r_b}\biggr]\,p_b^j\,\biggr\}\,.\label{C33}
\end{eqnarray}
The expectation value of this can then be written as
\begin{equation}\label{C34}
\delta_M B_{7d}= W_1+W_2\,,
\end{equation}
where
\begin{eqnarray}
W_1&=&\biggl\langle-\,\frac{Z^2}{8}\sum_{a,b,c\neq b}\frac{r_a^i}{r_a^3}\left(\frac{Zr_b^j}{r_b^3}
-\frac{r_{bc}^j}{r_{bc}^3}\right)\frac{r_b^ir_b^j-3\delta^{ij}r_b^2}{r_b}
-\frac{7\,Z^3}{4}\pi\delta^3(r_b)\biggr\rangle\nonumber\\
&=&\biggl\langle\frac{Z^3}{4r_1^3}+\frac{Z^3}{4r_2^3}+\frac{Z^3\,\vec{r}_1\cdot\vec{r}_2}{4r_1^3r_2^2}+\frac{Z^3\,\vec{r}_1\cdot\vec{r}_2}{4r_1^2r_2^3}
-\frac{7\,Z^3}{4}[\pi\delta^3(r_1)+\pi\delta^3(r_2)]\nonumber\\
&&+\,\frac{Z^2}{8}\sum_{b,c\neq b}\left(\frac{r_1^i}{r_1^3}+\frac{r_2^i}{r_2^3}\right)\frac{r_b^ir_b^j-3\delta^{ij}r_b^2}{r_b}\frac{r_{bc}^j}{r_{bc}^3}\biggr\rangle\,.\label{C35}
\end{eqnarray}
Here we used the identity (\ref{A26}) to rewrite the singular term as
\begin{equation}
\biggl\langle -\,Z\,\frac{r_b^ir_b^j-3\delta^{ij}r_b^2}{8\,r_b}\,\biggl(\nabla_b^i\,\frac{Z}{r_b}\biggr)\,\biggl(\nabla_b^j\,\frac{Z}{r_b}\biggr)\biggr\rangle=
\biggl\langle \frac14\,\frac{Z^3}{r_b^3}-\frac{7\,Z^3}{4}\,\pi\,\delta^3(r_b)\biggr\rangle\,.
\end{equation}
Further,
\begin{eqnarray}\label{C36}
W_2&=&\biggl\langle i\,\frac{Z^2}{16}\biggl(\,\sum_{a\neq b}\frac{r_a^i}{r_a^3}\,\biggl[p_b^2,\,\frac{r_b^ir_b^j-3\,\delta^{ij}r_b^2}{r_b}\biggr]\,p_b^j
+\sum_{b}\frac{r_b^i}{r_b^3}\,\biggl[p_b^2,\,\frac{r_b^ir_b^j-3\,\delta^{ij}r_b^2}{r_b}\biggr]p_b^j\biggr)\biggr\rangle\\
&=&\biggl\langle\frac{Z^2}{8}\,\sum_{a\neq b}p_b^k\,\frac{r_a^i}{r_a^3}\left(-\delta^{ik}\frac{r_b^j}{r_b}
+\delta^{jk}\frac{r_b^i}{r_b}-\delta^{ij}\frac{r_b^k}{r_b}-\frac{r_b^ir_b^jr_b^k}{r_b^3}\right)p_b^j\nonumber\\
&&+\,\frac{Z^2}{8r_1^4}+\frac{Z^2}{8r_2^4}+\frac{3\,Z^3}{4}[\pi\delta^3(r_1)+\pi\delta^3(r_2)]
+\frac{Z^2}{8}\sum_b p_b^j\,\frac{1}{r_b^4}\,\bigl(\delta^{jk}r_b^2-3r_b^jr_b^k\bigr)\,p_b^k\biggr\rangle\,,\nonumber
\end{eqnarray}
where for the reduction of the singular term we used the identity (\ref{A28}) from the Appendix, in particular
\begin{equation}
i\,\frac{Z^2}{16}\,\biggl\langle\frac{r_b^i}{r_b^3}\,\biggl[p_b^2,\,\frac{r_b^ir_b^j-3\,\delta^{ij}r_b^2}{r_b}\biggr]p_b^j\biggr\rangle
=\biggl\langle\frac18\,\frac{Z^2}{r_b^4}+\frac{3\,Z^3}{4}\,\pi\delta^3(r_b)
+\frac{Z^2}{8}\, p_b^j\,\frac{1}{r_b^4}\,\bigl(\delta^{jk}r_b^2-3r_b^jr_b^k\bigr)\,p_b^k\biggr\rangle\,.
\end{equation}

\section{Hydrogen limit}

In this section we perform the reduction of our general formulas to the hydrogenic limit for the
$S$ states, in order to demonstrate that the method reproduces the known results in agreement with
the Dirac equation and with the hydrogenic recoil corrections. First we treat the infinite nucleus
mass limit and then the recoil correction.

\subsection{Infinite nucleus mass limit}

We obtain the hydrogenic limit by sending ${r}_2\rightarrow\infty$ and consequently $\vec{p}_2\rightarrow 0$ and
${r}\rightarrow\infty$.
The effective operator $H^{(6)}$ reduces in the hydrogenic limit to (writing $r_1\equiv r$)
\begin{eqnarray}\label{D1}
H^{(6)} &=& \frac{p^6}{16}+\frac{1}{8}\,\bigl(\nabla V\bigr)^2 + \frac{5}{128}\,\bigl[p^2,\bigl[p^2,V\bigr]\bigr]-\frac{3}{32}\,p^2\,\nabla^2 V\,.
\end{eqnarray}

The first-order contribution to energy $B=\langle H^{(6)}\rangle$ is then
\begin{eqnarray}
B &=& \biggl\langle\,\frac{1}{4}\,(E-V)\,p^2\,(E-V) + \frac{1}{8}\,\frac{Z^2}{r^4}
+ \frac{5}{64}\,\bigl[E-V,\bigl[p^2,V\bigr]\bigr]-\frac{3}{16}\,(E-V)\,\nabla^2 V\biggr\rangle  \nonumber \\
          &=& \biggl\langle\,\frac{1}{8}\,\bigl[V,\bigl[p^2,V\bigr]\bigr] + \frac{1}{2}\,(E-V)^3
+ \frac{1}{8}\,\frac{Z^2}{r^4}-\frac{5}{32}\,\frac{Z^2}{r^4}-\frac{3\,E}{16}\, 4\pi Z\,\delta^3(r)\biggr\rangle \nonumber\\
%          &=& \left\langle\frac{7}{32} \frac{Z^2}{r^4} +\frac{1}{2}\bigl(-5 E^3 +3 E V^2 - V^3\bigr) - \frac{3E}{16}\, 4\pi Z\,\delta^3(r)\right\rangle \nonumber \\
          &=& \biggl\langle\,\frac{7}{32}\,\frac{Z^2}{r^4} - \frac{5}{2}\,E^3 + \frac{3\,E}{2}\,\frac{Z^2}{r^2}
+ \frac{1}{2}\,\frac{Z^3}{r^3} - \frac{3\,E}{16}\, 4\pi Z\,\delta^3(r) \biggr\rangle.\label{D2}
\end{eqnarray}

The operator $H^{(4)}$ reduces in the hydrogenic case to
\begin{eqnarray}\label{D3}
H^{(4)} = -\,\frac{p^4}{8} + \frac{Z\pi}{2}\,\delta^3(r),
\end{eqnarray}

which we again regularize as
\begin{eqnarray}\label{D4}
H^{(4)} = H_R + \{H-E,\,Q\},
\end{eqnarray}
with $Q\,=\, -Z/(4r) \, = \, V/4 $
and
\begin{eqnarray}\label{D5}
H_R = -\,\frac{1}{2}\,(E-V)^2 - \frac{Z}{4}\,\frac{\vec{r}\cdot\vec{\nabla}}{r^3}.
\end{eqnarray}
The second-order contribution to energy is then
\begin{eqnarray}
A &=& \biggl\langle H_R\,\frac{1}{(E-H)'}\,H_R\biggr\rangle + \langle Q\,(H-E)\,Q\rangle + 2\,\bigl\langle H^{(4)}\bigr\rangle \langle Q\rangle
- 2\,\bigl\langle Q \,H^{(4)}\bigr\rangle \nonumber \\
          &=& \biggl\langle H_R\,\frac{1}{(E-H)'}\,H_R\biggr\rangle
 + \frac{1}{32}\,\bigl\langle\bigl[V,\bigl[H-E,V\bigr]\bigr]\bigr\rangle + E\,E^{(4)} + \frac{1}{16}\,\bigl\langle V p^4\bigr\rangle \nonumber \\
%          &=& \left\langle H_R\,\frac{1}{(E-H)'}\,H_R\right\rangle
% + \left\langle \frac{1}{32}\frac{Z^2}{r^4} + E\,E^{(4)} + \frac{1}{16} \bigl[V,\bigl[ p^2 ,E-V\bigr]\bigr] + \frac{1}{4} V(E-V)^2 \right\rangle \nonumber \\
          &=& \biggl\langle H_R\,\frac{1}{(E-H)'}\,H_R\biggr\rangle
 + \biggl\langle-\,\frac{3}{32}\,\frac{Z^2}{r^4} + E\,E^{(4)} + \frac{E^3}{2} - \frac{E}{2}\,\frac{Z^2}{r^2} - \frac{1}{4}\,\frac{Z^3}{r^3} \biggr\rangle,\label{D6}
\end{eqnarray}
where we have used $\langle V \rangle = 2\,E$.
The sum of first- and second-order contributions is
\begin{eqnarray}\label{D7}
E^{(6)} &=& \biggl\langle H_R\,\frac{1}{(E-H)'}\,H_R\biggr\rangle \\
&& + \,\biggl\langle \,\frac{1}{8}\,\frac{Z^2}{r^4} + \frac{1}{4}\,\frac{Z^3}{r^3} - 2 \,E^3
+ E\,E^{(4)} + E\,\frac{Z^2}{r^2} - \frac{3\,E}{16}\, 4\pi Z\,\delta^3(r) \biggr\rangle\nonumber \\
        &=& \biggl\langle H_R\,\frac{1}{(E-H)'}\,H_R\biggr\rangle
 + \biggl\langle \,\frac{1}{8}\,\vec{p}\,\frac{Z^2}{r^2}\,\vec{p} + \frac{3\,E}{4}\,\frac{Z^2}{r^2} - 2\,E^3
+ E\,E^{(4)} - \frac{3\,E}{16}\, 4\pi Z\,\delta^3(r)\biggr\rangle\,,\nonumber
\end{eqnarray}
where we have used the indentity
\begin{eqnarray}\label{D8}
\frac{Z^2}{r^4} &=& \vec{p}\,\frac{Z^2}{r^2}\,\vec{p} - 2\left(E+\frac{Z}{r}\right)\frac{Z^2}{r^2}\,.
\end{eqnarray}

The expectation values of operators appearing in the final result are for S states
\begin{eqnarray}\label{hydr oprs}
E                 						&=&-\,\frac{Z^2}{2n^2},\label{D9} \\
E^{(4)}          	        				&=& \frac{3\,Z^4}{8n^4}-\frac{Z^4}{2n^3}, \label{D10}\\
\left\langle\frac{Z^2}{r^2}\right\rangle   			&=& \frac{2\,Z^4}{n^3}, \label{D11}\\
\left\langle\vec{p}\,\frac{Z^2}{r^2}\vec{p}\right\rangle 	&=&-\,\frac{2\,Z^6}{3n^5} + \frac{8Z^6}{3n^3},\label{D12} \\
\left\langle H_R\,\frac{1}{(E-H)'}\,H_R\right\rangle		&=&Z^6\left(-\,\frac{3}{8 n^6}+\frac{23}{24 n^5}-\frac{3}{8 n^4}-\frac{11}{24 n^3}\right),\label{D13}\\
\langle4\pi Z\,\delta^3 (r)\rangle				&=&\frac{4\,Z^4}{n^3}.\label{D14}
\end{eqnarray}

Substituting these values into energy we get the result
\begin{eqnarray}\label{D15}
E^{(6)} = Z^6\left(-\frac{5}{16n^6} + \frac{3}{4n^5} - \frac{3}{8n^4} - \frac{1}{8n^3}\right),
\end{eqnarray}
in agreement with the result from the Dirac equation obtained by expanding
\begin{eqnarray}\label{D16}
E_D = \left(1 + \frac{(Z\alpha)^2}{(n-1+\sqrt{1-(Z\alpha)^2})^2}\right)^{-\frac{1}{2}}
\end{eqnarray}
in the order $\alpha^6$.

\subsection{Recoil correction for hydrogenic limit}

Here the perturbation of the nonrelativistic Hamiltonian reduces to $\vec{P}^2/2M=\vec{p}\,{}^2/2M$.
This correction is then easily accounted for by making reduced mass rescaling $r\rightarrow\frac{r}{\mu}$ and expanding
the reduced mass as $(\mu/m)^n\approx 1-n\,\frac{m}{M}$.
The total recoil correction will then be the sum of the correction due to reduced mass
and the correction due to extra recoil operators in $H^{(6)}_M$ and $H^{(4)}_M$.
First we examine the reduced mass correction.

Rescaling the first-order operator $H^{(6)}$ and expanding up to the first order in nuclear mass
we obtain the recoil correction (utilizing results from the infinite nucleus mass limit)
\begin{eqnarray}
\delta_M B_1 &=& -5\,B + \left\langle -\,\frac{p^6}{16} + \frac{1}{8}\,\bigl(\nabla V\bigr)^2 \right\rangle \nonumber \\
%              &=& -5 B + \left\langle -\frac{1}{4} (E-V)\,p^2\,(E-V) + \frac{1}{8} \frac{Z^2}{r^4} \right\rangle \nonumber \\
%	      &=& -5 B + \left\langle -\frac{1}{8} \frac{Z^2}{r^4} - \frac{1}{2}(E-V)^3 \right\rangle \nonumber \\
	      &=& -5\,B + \left\langle -\,\frac{1}{8}\,\frac{Z^2}{r^4} + \frac{5\,E^3}{2} - \frac{3\,E}{2}\,\frac{Z^2}{r^2}-\frac{1}{2}\,\frac{Z^3}{r^3} \right\rangle.\label{D17}
\end{eqnarray}

The second-order contribution due to reduced mass is
\begin{eqnarray}\label{D18}
\delta_M A_1 &=& -5\,A +  \left\langle \frac{p^4}{4}\,\frac{1}{(E-H)'}\,H^{(4)}\right\rangle \\
              &=& -5\,A + \left\langle (E-V)^2\,\frac{1}{(E-H)'}\,H^{(4)}\right\rangle
+ \left\langle (E-V)\,(H-E)\,\frac{1}{(E-H)'}\, H^{(4)}\right\rangle \nonumber \\
%	      &=& -5 A + \left\langle (E-V)^2\,\frac{1}{(E-H)'}\,H_R\right\rangle + \left\langle (E-V)^2\,\frac{1}{(E-H)'}\,(H-E)\,Q\right\rangle \nonumber \\
%     && + \,\langle E-V\rangle\,E^{(4)} - \bigl\langle (E-V)\,H^{(4)}\bigr\rangle\nonumber\\
%	      &=& -5 A + \delta_M A_1' + \left\langle \frac{E}{2}\,(E-V)^2 - \frac{1}{4}(E-V)^2\,V - 2 E\,E^{(4)} + V H^{(4)} \right\rangle \nonumber \\
%	      &=& -5 A + \delta_M A_1' + \left\langle - 2 E\,E^{(4)} - 2E^3 + E\,V^2 - \frac{1}{4}V^3 - \frac{1}{4}V\,p^2\,(E-V) \right\rangle \nonumber \\
%	      &=& -5 A + \delta_M A_1' + \left\langle - 2 E\,E^{(4)} - 2E^3 + E\,V^2 - \frac{1}{4}V^3 - \frac{1}{8}\left[V,\left[p^2,E-V\right]\right]-\frac{1}{2}V(E-V)^2\right\rangle \nonumber\\
	      &=& -5\,A + \left\langle (E-V)^2\,\frac{1}{(E-H)'}\,H_R\right\rangle
 + \left\langle - \,2 \,E\,E^{(4)} - 3\,E^3 + 2\,E\,\frac{Z^2}{r^2} + \frac{3}{4}\,\frac{Z^3}{r^3} + \frac{1}{4}\,\frac{Z^2}{r^4} \right\rangle\,.\nonumber
\end{eqnarray}

Summing now both terms $\delta_M A_1$ and $\delta_M B_1$ we get the total
recoil correction due to the reduced mass rescaling $E_\textrm{i}$,
\begin{eqnarray}\label{D19}
E_\textrm{i} &=& \delta_M A_1+\delta_M B_1\\
%             &=& -5 E^{(6)} + \left\langle (E-V)^2\,\frac{1}{(E-H)'}\,H_R\right\rangle
% + \left\langle -\frac{E^3}{2} - 2 E\,E^{(4)} + \frac{E}{2}\frac{Z^2}{r^2} + \frac{1}{4}\frac{Z^3}{r^3} + \frac{1}{8}\frac{Z^2}{r^4} \right\rangle \nonumber \\
             &=& -5\,E^{(6)} + \left\langle (E-V)^2\,\frac{1}{(E-H)'}\,H_R\right\rangle
 + \left\langle -\,\frac{E^3}{2} - 2 \,E\,E^{(4)} + \frac{E}{4}\,\frac{Z^2}{r^2} + \frac{1}{8}\,\vec{p}\,\frac{Z^2}{r^2}\,\vec{p} \right\rangle,\nonumber
\end{eqnarray}
where we have again used the identity (\ref{D8}).

The next contribution comes from the extra recoil operators. The recoil correction to the Breit Hamiltonian $H^{(4)}$ is
\begin{eqnarray}\label{D20}
\delta_M H^{(4)} = -\frac{Z}{2} \,p^i\left(\frac{\delta^{ij}}{r}+\frac{r^i r^j}{r^3}\right)P^j,
\end{eqnarray}
and the corresponding second-order correction to energy is
\begin{eqnarray}\label{D21}
\delta_M A_2 &=& -\,Z\,\biggl\langle p^i\left(\frac{\delta^{ij}}{r}+\frac{r^i r^j}{r^3}\right)p^j\,\frac{1}{(E-H)'}\,H^{(4)} \biggr\rangle \\
%             &=& -Z\left\langle p^i\left(\frac{\delta^{ij}}{r}+\frac{r^i r^j}{r^3}\right) p^j\,\frac{1}{(E-H)'}\,H_R \right\rangle
%- Z \left\langle p^i\left(\frac{\delta^{ij}}{r}+\frac{r^i r^j}{r^3}\right) p^j\,\frac{1}{(E-H)'}\,(H-E)\,Q\right\rangle \nonumber \\
%	     &=& \delta_M A_2' - \frac{Z}{4} \left\langle p^i\left(\frac{\delta^{ij}}{r}+\frac{r^i r^j}{r^3}\right) p^j\right\rangle \langle V\rangle
%+ \frac{Z}{4} \left\langle p^i\left(\frac{\delta^{ij}}{r}+\frac{r^i r^j}{r^3}\right) p^j\,V \right\rangle \nonumber \\
%	     &=& \delta_M A_2' - \frac{Z}{2} E\left\langle p^i\left(\frac{\delta^{ij}}{r}+\frac{r^i r^j}{r^3}\right) p^j\right\rangle
%- \frac{Z^2}{4} \left\langle p^i\left(\frac{\delta^{ij}}{r^2}+\frac{r^i r^j}{r^4}\right) p^j \right\rangle \nonumber\\
%&&- \frac{Z^2}{8} \left\langle\left[p^i,\left[p^j,\frac{1}{r}\right]\right]\left(\frac{\delta^{ij}}{r}+\frac{r^i r^j}{r^3}\right)\right\rangle\nonumber \\
	     &=& -\,Z\,\biggl\langle p^i\left(\frac{\delta^{ij}}{r}+\frac{r^i r^j}{r^3}\right) p^j\,\frac{1}{(E-H)'}\,H_R \biggr\rangle\nonumber\\
&& +\,\biggl\langle-\,\frac{E Z}{2} p^i\left(\frac{\delta^{ij}}{r}+\frac{r^i r^j}{r^3}\right) p^j
- \frac{Z^2}{4} \,p^i\left(\frac{\delta^{ij}}{r^2}+\frac{r^i r^j}{r^4}\right) p^j + \frac{1}{4}\,\frac{Z^2}{r^4} + \frac{Z^3}{2}\pi\,\delta^3(r)\biggr\rangle.\nonumber
\end{eqnarray}

The correction due to the extra first-order recoil operators is
\begin{eqnarray}
\delta_M B_2 &=& \langle\, \delta_M H^{(6)}\,\rangle \nonumber\\
&=&\biggl\langle\frac{1}{4}\,\frac{Z^2}{r^4} + \frac{Z^3}{2}\pi\,\delta^3(r) + \frac{Z}{4} \,p^2\,p^i\left(\frac{\delta^{ij}}{r}+\frac{r^i r^j}{r^3}\right) p^j
+ \frac{Z^2}{8} \,p^i\left(\frac{\delta^{ij}}{r^2}+3\frac{r^i r^j}{r^4}\right) p^j\nonumber \\
&&+ \,i\,\frac{Z^2}{8} \frac{r^i}{r^3}\left[H-E,\,\frac{r^i r^j - 3\delta^{ij}\,r^2}{r}\,p^j\right]\biggr\rangle\nonumber\\
&=& \delta_M B_{2a}+\delta_M B_{2b}+\delta_M B_{2c}+\delta_M B_{2d}+\delta_M B_{2e}.\label{D22}
\end{eqnarray}
The third and the fifth terms are
\begin{eqnarray}
\delta_M B_{2c} &=& \biggl\langle\,\frac{Z}{2}\,(E-V)\,p^i\left(\frac{\delta^{ij}}{r}+\frac{r^i r^j}{r^3}\right) p^j\biggr\rangle \nonumber \\
  &=& \biggl\langle\,\frac{Z}{2} \,p^i \,(E-V)\left(\frac{\delta^{ij}}{r}+\frac{r^i r^j}{r^3}\right) p^j
 + \frac{Z^2}{4} \left[p^i,\left[p^j,\frac{1}{r}\right]\right] \left(\frac{\delta^{ij}}{r}+\frac{r^i r^j}{r^3}\right)\biggr\rangle\nonumber\\
 &=& \biggl\langle\,\frac{Z}{2} \,p^i \left(E+\frac{Z}{r}\right)\left(\frac{\delta^{ij}}{r}+\frac{r^i r^j}{r^3}\right) p^j
- \frac{1}{2}\,\frac{Z^2}{r^4} - Z^3\pi\,\delta^3(r)\biggr\rangle\,,\label{D23}
\end{eqnarray}
and
\begin{eqnarray}
\delta_M B_{2e} &=&\biggl\langle\,i\,\frac{Z^2}{8}\frac{r^i}{r^3}\left\{ \bigl[V,\,p^j\bigr]\,\frac{r^i r^j - 3\delta^{ij}\,r^2}{r}
+ \left[\frac{p^2}{2},\,\frac{r^i r^j - 3\delta^{ij}\,r^2}{r}\right]p^j\right\}\biggr\rangle\nonumber\\
  &=& \biggl\langle\,\frac{1}{4}\,\frac{Z^3}{r^3} + \frac{1}{8}\,\frac{Z^2}{r^4} - Z^3\pi\,\delta^3(r)
+ \frac{Z^2}{8} \,p^i\,\frac{1}{r^4}\left(\delta^{ij}\,r^2 - 3r^i r^j\right) p^j\biggr\rangle\,.\label{D24}
\end{eqnarray}
The first-order contribution $\delta_M B_2$ is the sum of all terms $\delta_M B_{2a}\ldots\delta_M B_{2e}$ and is
\begin{eqnarray}
\delta_M B_2&=& \biggl\langle-\,\frac{1}{8}\,\frac{Z^2}{r^4} + \frac{1}{4}\,\frac{Z^3}{r^3} - \frac{3Z^3}{2}\pi\,\delta^3(r)
+ \frac{Z}{2} \,p^i \left(E+\frac{Z}{r}\right)\left(\frac{\delta^{ij}}{r}+\frac{r^i r^j}{r^3}\right) p^j
+ \frac{1}{4} \,\vec{p}\,\frac{Z^2}{r^2}\,\vec{p}\biggr\rangle\,.\nonumber\\\label{D25}
\end{eqnarray}
The correction $E_\textrm{ii}$ due to extra recoil operators is then the sum of $\delta_M B_2$
and $\delta_M A_2$,
\begin{eqnarray}
E_\textrm{ii} &=& \delta_M A_2+\delta_M B_2 \nonumber\\
% &=& -Z\left\langle p^i\left(\frac{\delta^{ij}}{r}+\frac{r^i r^j}{r^3}\right) p^j\,\frac{1}{(E-H)'}\,H_R \right\rangle\nonumber\\
%&& + \left\langle \frac{Z^2}{4} p^i \left(\frac{\delta^{ij}}{r^2}+\frac{r^i r^j}{r^4}\right) p^j + \frac{1}{8}\frac{Z^2}{r^4} + \frac{1}{4}\frac{Z^3}{r^3}
%- Z^3\pi\,\delta^3(r) + \frac{1}{4}\vec{p}\,\frac{1}{r^2}\,\vec{p}\right\rangle \nonumber \\
&=& -\,Z\,\biggl\langle p^i\left(\frac{\delta^{ij}}{r}+\frac{r^i r^j}{r^3}\right) p^j\,\frac{1}{(E-H)'}\,H_R \biggr\rangle\nonumber\\
&& +\,\biggl\langle \frac{Z^2}{4} \,p^i \left(\frac{\delta^{ij}}{r^2}+\frac{r^i r^j}{r^4}\right) p^j - Z^3\pi\,\delta^3(r)
+ \frac{3}{8}\,\vec{p}\,\frac{1}{r^2}\,\vec{p} -\frac{E}{4}\,\frac{Z^2}{r^2}\biggr\rangle.\label{D26}
\end{eqnarray}

Finally, the total recoil correction for the S state hydrogenic limit
is the sum of the reduced mass scaling correction $E_\textrm i$ and the correction due to extra operators $E_\textrm {ii}$ and is
\begin{eqnarray}\label{D27}
\delta_M E^{(6)} &=& E_\textrm{i}+E_\textrm{ii}\\
 &=& \biggl\langle (E-V)^2\,\frac{1}{(E-H)'}\,H_R\biggr\rangle
-Z\,\biggl\langle p^i\left(\frac{\delta^{ij}}{r}+\frac{r^i r^j}{r^3}\right) p^j\,\frac{1}{(E-H)'}\,H_R \biggr\rangle\nonumber\\
&& -\, 5 \,E^{(6)} - \frac{E^3}{2} - 2 \,E\,E^{(4)} +\,\biggl\langle \,\frac{1}{2}\,\vec{p}\,\frac{Z^2}{r^2}\,\vec{p}
+\frac{Z^2}{4} p^i \left(\frac{\delta^{ij}}{r^2}+\frac{r^i r^j}{r^4}\right) p^j -Z^3 \pi\,\delta^3(r)\biggr\rangle\,.\nonumber
\end{eqnarray}

In addition to the operators already used, the expectation values are
\begin{eqnarray}
\left\langle Z^2 \,p^i \left(\frac{\delta^{ij}}{r^2}+\frac{r^i r^j}{r^4}\right) p^j\right\rangle
        &=& -\,\frac{4\,Z^6}{3n^5} + \frac{16\,Z^6}{3n^3},\label{D28} \\
\biggl\langle (E-V)^2\,\frac{1}{(E-H)'}\,H_R\biggr\rangle
 &=& Z^6\left(\frac{1}{2 n^6}-\frac{9}{4 n^5}+\frac{3}{2 n^4}+\frac{3}{2 n^3}\right), \label{D29}\\
Z\,\biggl\langle p^i\left(\frac{\delta^{ij}}{r}+\frac{r^i r^j}{r^3}\right) p^j\,\frac{1}{(E-H)'}\,H_R \biggr\rangle
&=& Z^6\left(\frac{2}{n^6}-\frac{37}{6 n^5}+\frac{3}{n^4}+\frac{11}{3 n^3} \right).\label{D30}
\end{eqnarray}
Using these expectation values the final result is
\begin{eqnarray}\label{D31}
\delta_M E^{(6)} = Z^6\left(\frac{1}{2 n^6}-\frac{1}{n^5}+\frac{3}{8 n^4}+\frac{1}{8 n^3}\right),
\end{eqnarray}
in agreement with the result from the Dirac equation
\begin{eqnarray}\label{D32}
E^M_D = \frac{1-E_D^2}{2}
\end{eqnarray}
expanded in the order $\alpha^6$. In particular, it vanishes for the hydrogenic ground state.

%Tables%

\begin{table*}
\caption{Expectation values of operators $Q_i$ and the corresponding recoil corrections, with $i = 1\ldots 30$.}
\scriptsize

\label{oprsQ}
\begin{ruledtabular}
\begin{tabular}{ldddd}
 &   \multicolumn{2}{c}{2$^3S$}&   \multicolumn{2}{c}{2$^3P$}\\
 & \multicolumn{1}{c}{$\langle Q_i\rangle$} & \multicolumn{1}{c}{$\delta_M \langle Q_i \rangle$} & \multicolumn{1}{c}{$\langle Q_i\rangle$} & \multicolumn{1}{c}{$\delta_M \langle Q_i \rangle$}\\ \hline
$Q_1 =4 \pi \delta^3 (r_1)$   				     &  16.592\,071   & -49.748\,907        & 15.819\,309 &-48.358\,598 \\
$Q_2 =4 \pi \delta^3 (r)$               	             &   0 	    &   0 	 &  0&  0\\	
$Q_3 =4 \pi \delta^3(r_1)/r_2$                  	     &   4.648\,724   & -18.821\,266 &  4.349\,766 &-14.576\,147\\	
$Q_4 =4 \pi \delta^3(r_1)\, p_2^2$ 	                     &   2.095\,714   & -10.638\,077 &  4.792\,830 &-17.366\,064\\
$Q_5 =4 \pi \delta^3(r)/r_1$				     &   0	    &   0       &  0 & 0\\
$Q_6 =4 \pi\,\vec{p}\,\delta^3(r)\,\vec{p} $		     &   0.028\,099   &  -0.163\,026 &  0.077\,524 & -0.100\,949\\
$Q_7 =1/r$						     &   0.268\,198   &  -0.272\,645        &  0.266\,641 & -0.082\,865\\
$Q_8 =1/r^2$						     &   0.088\,906   &  -0.182\,363    &  0.094\,057 & -0.052\,275\\
$Q_9 =1/r^3$                    	                     &   0.038\,861   &  -0.121\,355 &  0.047\,927 & -0.036\,603\\
$Q_{10}=1/r^4$                  	                     &   0.026\,567   &  -0.113\,712 &  0.043\,348 & -0.042\,669\\
$Q_{11}=1/r_1^2$                	                     &   4.170\,446   &  -8.338\,455 &  4.014\,865 & -8.127\,584\\
$Q_{12}=1/(r_1 r_2)$            	                     &   0.560\,730   &  -1.147\,101 &  0.550\,342 & -0.709\,019\\
$Q_{13}=1/(r_1 r)$              	                     &   0.322\,696   &  -0.657\,458 &  0.317\,639 & -0.381\,158\\
$Q_{14}=1/(r_1 r_2 r)$          	                     &   0.186\,586   &  -0.576\,097 &  0.198\,346 & -0.295\,115\\
$Q_{15}=1/(r_1^2 r_2)$					     &   1.242\,704   &  -3.791\,743 &  1.196\,631 & -2.687\,288\\
$Q_{16}=1/(r_1^2 r)$					     &   1.164\,599   &  -3.545\,640 &  1.109\,463 & -2.554\,378\\
$Q_{17}=1/(r_1 r^2)$   					     &   0.112\,360   &  -0.346\,820 &  0.121\,112 & -0.166\,459\\
$Q_{18}=(\vec{r}_1\cdot\vec r)/(r_1^3 r^3)$                  &   0.011\,331   &  -0.055\,997 &  0.030\,284 & -0.030\,290\\
$Q_{19}=(\vec{r}_1\cdot\vec r)/(r_1^3 r^2)$                  &   0.054\,635   &  -0.211\,280 &  0.075\,373 & -0.104\,553\\
$Q_{20}=r_1^i r_2^j(r^i r^j-3\delta^{ij}r^2)/(r_1^3 r_2^3 r)$&   0.027\,082   &  -0.256\,024 &  0.090\,381 & -0.166\,239\\
$Q_{21}=p_2^2/r_1^2$					     &   0.751\,913   &  -3.075\,881 &  1.410\,228 & -3.635\,740\\
$Q_{22}=\vec{p}_1/r_1^2\, \vec{p}_1$			     &  16.720\,479   & -66.901\,955 & 15.925\,672 &-64.131\,339\\
$Q_{23}=\vec{p}_1/r^2\, \vec{p}_1$			     &   0.243\,754   &  -1.008\,306 &  0.279\,229 & -0.572\,398\\
$Q_{24}=p_1^i\,(r^i r^j+\delta^{ij} r^2)/(r_1 r^3)\, p_2^j$  &   0.002\,750   &  -0.068\,255 & -0.097\,364 & -0.056\,872\\
$Q_{25}=P^i\, (3 r^i r^j-\delta^{ij} r^2)/r^5\, P^j$	     &   0.062\,031   &  -0.336\,586 & -0.060\,473 &  0.119\,687\\
$Q_{26}=p_2^k \,r_1^i\,/r_1^3 (\delta^{jk} r^i/r - \delta^{ik} r^j/r -\delta^{ij} r^k/r -r^i r^j
r^k/r^3)\, p_2^j$		     &  -0.009\,102   &   0.035\,209 &  0.071\,600 & -0.134\,238\\
$Q_{27}=p_1^2\, p_2^2$					     &   0.488\,198   &  -1.988\,286 &  1.198\,492 & -3.171\,122\\
$Q_{28}=p_1^2\,/r_1\, p_2^2$				     &   1.597\,727   &  -8.106\,766 &  3.883\,405 &-13.814\,978\\
$Q_{29}=\vec{p}_1\times\vec{p}_2\,/r\,\vec{p}_1\times\vec{p}_2$
							     &   0.070\,535   &  -0.358\,089 &  0.399\,306 & -1.076\,373\\
$Q_{30}=p_1^k \,p_2^l\,(-\delta^{jl} r^i r^k/r^3 - \delta^{ik} r^j r^l/r^3 +3r^i r^j r^k r^l/r^5)\,
p_1^i\, p_2^j$			     &  -0.034\,780   &   0.177\,968 & -0.187\,305 &  0.490\,555\\
\end{tabular}
\end{ruledtabular}
\end{table*}

\begin{table*}
\caption{Expectation values of operators $Q_i$ with $i = 31\ldots 50$, the expectation value of the Breit Hamiltonian $E^{(4)}$ and the first-order
recoil corrections $\delta_M E$ and $\delta_M E^{(4)}$.}
\scriptsize

\label{oprsQ2}
\begin{ruledtabular}
\begin{tabular}{lddd}
   & 2^3S & 2^3P  \\ \hline
$Q_{31}=4 \pi \delta^3(r_1)\, \vec{p}_1\cdot\vec{p}_2$       &   0.040\,294 & -0.457\,224   \\
$Q_{32}=(\vec{r}_1\cdot\vec{r}_2)/(r_1^3 r_2^3)$             &  -0.005\,797 & -0.032\,383   \\
$Q_{33}=\vec{p}_1\cdot\vec{p}_2$			     &   0.007\,442 & -0.064\,572  \\
$Q_{34}=\vec{P}\,/r_1\,\vec{P}$				     &   4.974\,707 &  4.730\,359  \\
$Q_{35}=\vec{P}\,/r\,\vec{P}$				     &   1.232\,372 &  1.127\,146   \\
$Q_{36}=\vec{P}\,/r_1^2\,\vec{P}$                            &  17.504\,835 & 16.972\,775   \\
$Q_{37}=\vec{P}\,/(r_1 r_2)\,\vec{P}$			     &   2.489\,592 &  2.291\,176   \\
$Q_{38}=\vec{P}\,/(r_1 r)\,\vec{P}$			     &   1.454\,007 &  1.350\,214   \\
$Q_{39}=\vec{P}\,/r^2\,\vec{P}$				     &   0.438\,804 &  0.413\,144   \\
$Q_{40}=p_1^2\,p_2^2\,P^2$				     &  10.324\,509 & 24.527\,699   \\
$Q_{41}=P^2\,p_1^i\, (r^i r^j + \delta^{ij} r^2)/r^3 \, p_2^j$
							     &   0.151\,748 &  0.067\,201   \\
$Q_{42}=p_1^i\,(r_1^i r_1^j + \delta^{ij} r_1^2)/r_1^4 \,  P^j$
						             & 33.461\,709  & 31.489\,835  \\
$Q_{43}=p_1^i\,(r_1^i r_1^j + \delta^{ij} r_1^2)/(r_1^3 r_2)\,  P^j$
						             &  2.486\,269  & 2.217\,310  \\
$Q_{44}=p_1^i\,p_2^k\,(r_1^ir_1^j+\delta^{ij}r_1^2)/r_1^3\,p_2^k\, P^j$
							     &  1.100\,915  & 2.527\,505   \\
$Q_{45}=p_2^i(r^i r^j+\delta^{ij} r^2)(r_1^jr_1^k
+\delta^{jk} r_1^2)/(r_1^3 r^3)\, P^k$	 	             &  0.540\,877  & 0.467\,623  \\
$Q_{46}=p_1^i(r_1^i r_1^j+\delta^{ij} r_1^2)(r_2^jr_2^k
+\delta^{jk} r_2^2)/(r_1^3 r_2^3)\, p_2^k$		     &   0.006\,782 & -0.201\,826   \\
$Q_{47}=(\vec{r}_1\cdot\vec{r}_2)/(r_1^3 r_2^2)$             &  -0.008\,117 & -0.028\,621   \\
$Q_{48}=r_1^i r^j(r_1^i r_1^j-3\delta^{ij} r_1^2)/(r_1^4r^3)$&  -0.036\,861 & -0.057\,404   \\
$Q_{49}=r_1^ir^j(r_2^ir_2^j-3\delta^{ij}r_2^2)/(r_1^3r_2r^3)$&  -0.089\,086 & -0.126\,780   \\
$Q_{50}=p_2^k\,r_1^i/r_1^3\,(\delta^{jk}r_2^i/r_2-\delta^{ik}r_2^j/r_2
-\delta^{ij}r_2^k/r_2-r_2^ir_2^jr_2^k/r_2^3)\,p_2^j$	     &   0.005\,856 & -0.092\,036   \\
$ E^{(4)}$ & -2.164\,477\,972 & -1.967\,358\,377 \\
$\delta_M E $ & 2.182\,671\,509 & 2.068\,591\,766\\
$\delta_M E^{(4)}$  & 0.089\,185\,018 & 0.230\,100\,830 \\
\end{tabular}
\end{ruledtabular}
\end{table*}

\begin{table*}
\caption{Individual $\alpha^6\,m^2/M$ recoil corrections to the ionization energies of  the $2^3S$ and $2^3P$ states.}
\label{Es}
\begin{ruledtabular}
\begin{tabular}{ldd}
Term & 2^3S & 2^3P \\ \hline
$E_\textrm{i}$     &  1.190\,05 &  0.853\,52(10)\\
$E_\textrm{ii}$    &  0.044\,46 &  0.032\,33(5)\\
$E_\textrm{iii}$   &  0.025\,11 &  1.197\,09(10)\\
$E_\textrm{iv}$    &  0.018\,48 &  0.013\,00(6)\\
$E_\textrm{v}$     &-58.048\,06 &-52.977\,21\\
$E_\textrm{vi}$    & 56.945\,69 & 52.339\,33\\ \hline
Subtotal           &0.175\,72&1.458\,05(16)\\
$E_\textrm{vii}$   &-11.867\,15 & -1.891\,19 \\ \hline
Sum        &-11.691\,45 & -0.433\,13(16) \\ \hline
$\delta_M E^{(6)}(\textrm{kHz})$& -29.91 & -1.11 \\
\end{tabular}
\end{ruledtabular}
\end{table*}

%\begin{table}[!htb]
%\renewcommand{\arraystretch}{0.8}
%\caption{Contributions of the order $\alpha^m\,(m_e/M)^n\,m_e c^2/h$ to $(2^3P$-$2^3S)_{\rm centroid}$ transition energy in MHz. FNS stands for
%the contribution due to finite size of the nucleus.}
%\label{TBL3}
%\begin{center}
%\begin{tabular}{c | w{10.6}w{6.6}w{6.6}w{10.6}}
%\hline
%\hline
%   & \centt{$(m_e/M)^0$}  & \centt{$(m_e/M)^1$}  & \centt{$(m_e/M)^2$}   &  \centt{Sum} \\
%\hline
%%
% $\alpha^2$        &      276\,775\,637.53        &  -102\,903.45    &            4.78        &      276\,672\,738.86\\
% $\alpha^4$        &          69\,066.19        &        6.77    &              \textrm{---}      &                69\,072.96 \\
% $\alpha^5$        &          -5\,234.16        &        0.19    &              \textrm{---}       &               -5\,233.97 \\
% $\alpha^6$        &            -87.06        &        0.03     &              \textrm{---}       &              -87.03 \\
% $\alpha^7$        &              8.03 (1.00) &       -0.01(1)       &              \textrm{---}        &                 8.02 (1.00) \\
% FNS                  &             -3.45        &         \textrm{---}      &              \textrm{---}      &                  -3.45 \\
%\hline
% $E_{\rm th}$\cite{yerokhin:10:helike}         &&&&     276\,736\,495.39 (1.00)\\
% $E_{\rm exp}$\cite{cancio}        &&&&     276\,736\,495.649(2) \\
% \hline \hline
%\end{tabular}
%\end{center}
%%\end{ruledtabular}
%\end{table}

\begin{table*}
\caption{Breakdown of theoretical contributions to the $2^3S$--$2^3P$ centroid transition energy for ${}^4\textrm{He}$, in MHz.
The uncertainty due to approximate $\alpha^7$ contribution is assumed to be 1 MHz, i.e. four times less than in our previous work \cite{yerokhin:10:helike}.
FNS is a finite nuclear size and NPOL the nuclear polarizability corrections. }
\label{TBL3}
\begin{center}
\begin{tabular}{c | w{12.6}w{6.6}w{2.6}w{10.6}}
\hline
\hline
   & \centt{$(m/M)^0$}  & \centt{$(m/M)^1$}  & \centt{$(m/M)^2$}   &  \centt{Sum} \\
\hline
 $\alpha^2$        &      -276\,775\,637.536        &  102\,903.459         &     -4.781        &      -276\,672\,738.857\\
 $\alpha^4$        &            -69\,066.189       &         -6.769         &     -0.003    &               -69\,072.961 \\
 $\alpha^5$        &              5\,234.163        &        -0.186         & \textrm{---}     &              5\,233.978 \\
 $\alpha^6$        &                 87.067       &          -0.029         & \textrm{---}     &                  87.039 \\
 $\alpha^7$        &                 -8.0\,(1.0)  &         \textrm{---}    & \textrm{---}     &                 -8.0 (1.0) \\
 FNS               &                  3.427         &         \textrm{---}  & \textrm{---}     &                  3.427 \\
 NPOL              &                 -0.002         &         \textrm{---}  & \textrm{---}     &                 -0.002  \\
\hline
Present theory                            &&&&      -276\,736\,495.41\,(1.00) \\
Previous theory \cite{yerokhin:10:helike}                    &&&&      -276\,736\,495.37\,(4.00) \\
Exp. \cite{cancio} + Th. ${}^3P_{0}$--${}^3P_{2}$ \cite{hefs}
                                        &&&&      -276\,736\,495.649\,(2) \\
 \hline \hline
\end{tabular}
\end{center}
\end{table*}

\begin{table*}
\caption{Breakdown of theoretical contributions to the ${}^3\textrm{He}-{}^4\textrm{He}$ isotope shift of
the $2^3S$--$2^3P$ centroid transition energy, for the point nucleus, in kHz. EMIX is an additional correction
in $^3$He due to the second-order hyperfine singlet-triplet mixing \cite{pachucki:15:heis}.}
\label{TBL5}
\begin{center}
\begin{tabular}{c | w{10.6}w{6.6}w{6.6}w{10.6}}
\hline
\hline
   & \centt{$(m/M)^1$}  & \centt{$(m/M)^2$} & \centt{$(m/M)^3$}   &  \centt{Sum} \\
\hline
 $\alpha^2$        &  33\,673\,018.7  & -3\,640.6 & 0.4               & 33\,669\,378.5 \\
 $\alpha^4$        &     -2\,214.9  &    -2.4 & \textrm{---}      &    -2\,217.3 \\
 $\alpha^5$        &        -60.7 & \textrm{---} & \textrm{---} &      -60.7   \\
 $\alpha^6$        &         -9.4 & \textrm{---} & \textrm{---} &       -9.4      \\
 $\alpha^7$        &          0.0\,(0.9) & \textrm{---} & \textrm{---} &   0.0\,(0.9)    \\
 NPOL              &         -1.1 &\textrm{---} & \textrm{---}  &       -1.1            \\
 EMIX              &         \textrm{---} & 54.6 & \textrm{---} &        54.6               \\
\hline
Present theory     &&&& 33\,667\,149.3(0.9)\\
 \hline \hline
\end{tabular}
\end{center}
\end{table*}

\end{document}